\documentclass[a4paper,11pt]{article}
\pdfoutput=1 

\usepackage{jinstpub} 
\newcommand\numberthis{\addtocounter{equation}{1}\tag{\theequation}}

\title{Application and performance of an ML-EM algorithm in NEXT.}

\author[a,1]{A. Sim\'on, \note{Corresponding author.}}
\author[b]{C. Lerche,}
\author[m]{F.~Monrabal,}
\author[a,2]{J.J.~G\'omez-Cadenas,\note{NEXT Co-spokesperson.}}

\author[a]{V.~\'Alvarez,}
\author[c]{C.D.R.~Azevedo,}
\author[a]{J.M.~Benlloch-Rodr\'{i}guez,}
\author[d]{F.I.G.M.~Borges,}
\author[a]{A.~Botas,}
\author[a]{S.~C\'arcel,}
\author[a]{J.V.~Carri\'on,}
\author[e]{S.~Cebri\'an,}
\author[d]{C.A.N.~Conde,}
\author[a]{J.~D\'iaz,}
\author[f]{M.~Diesburg,}
\author[d]{J.~Escada,}
\author[g]{R.~Esteve,}
\author[a]{R.~Felkai,}
\author[h]{L.M.P.~Fernandes,}
\author[a]{P.~Ferrario,}
\author[c]{A.L.~Ferreira,}
\author[h]{E.D.C.~Freitas,}
\author[i]{A.~Goldschmidt,}
\author[j]{D.~Gonz\'alez-D\'iaz,}
\author[k]{R.M.~Guti\'errez,}
\author[l]{J.~Hauptman,}
\author[h]{C.A.O.~Henriques,}
\author[k]{A.I.~Hernandez,}
\author[j]{J.A.~Hernando~Morata,}
\author[g]{V.~Herrero,}
\author[m]{B.J.P.~Jones,}
\author[n]{L.~Labarga,}
\author[a]{A.~Laing,}
\author[f]{P.~Lebrun,}
\author[a]{I.~Liubarsky,}
\author[a]{N.~L\'opez-March,}
\author[k]{M.~Losada,}
\author[a,3]{J.~Mart\'in-Albo,\note{Now at University of Oxford, United Kingdom.}}
\author[j]{G.~Mart\'inez-Lema,}
\author[a]{A.~Mart\'inez,}
\author[m]{A.D.~McDonald,}
\author[h]{C.M.B.~Monteiro,}
\author[g]{F.J.~Mora,}
\author[c]{L.M.~Moutinho,}
\author[a]{J.~Mu\~noz Vidal,}
\author[a]{M.~Musti,}
\author[a]{M.~Nebot-Guinot,}
\author[a]{P.~Novella,}
\author[m,2]{D.R.~Nygren}
\author[a]{B.~Palmeiro,}
\author[f]{A.~Para,}
\author[a]{J.~P\'{e}rez,}
\author[a]{M.~Querol,}
\author[a]{J.~Renner,}
\author[o]{L.~Ripoll,}
\author[a]{J.~Rodr\'iguez,}
\author[m]{L.~Rogers,}
\author[d]{F.P.~Santos,}
\author[h]{J.M.F.~dos~Santos,}
\author[p,4]{C.~Sofka,\note{Now at University of Texas at Austin, USA.}}
\author[a]{M.~Sorel,}
\author[p]{T.~Stiegler,}
\author[g]{J.F.~Toledo,}
\author[a]{J.~Torrent,}
\author[q]{Z.~Tsamalaidze,}
\author[c]{J.F.C.A.~Veloso,}
\author[p]{R.~Webb,}
\author[p,5]{J.T.~White,\note{Deceased.}}
\author[a]{N.~Yahlali}
\affiliation[a]{
Instituto de F\'isica Corpuscular (IFIC), CSIC \& Universitat de Val\`encia\\
Calle Catedr\'atico Jos\'e Beltr\'an, 2, 46980 Paterna, Valencia, Spain}
\affiliation[b]{Institute of Neuroscience and Medicine (INM-4), Forschungszentrum J\" ulich GmbH\\
 J\" ulich, Germany}
\affiliation[c]{
Institute of Nanostructures, Nanomodelling and Nanofabrication (i3N), Universidade de Aveiro\\
Campus de Santiago, 3810-193 Aveiro, Portugal}
\affiliation[d]{
LIP, Department of Physics, University of Coimbra\\
P-3004 516 Coimbra, Portugal}
\affiliation[e]{
Laboratorio de F\'isica Nuclear y Astropart\'iculas, Universidad de Zaragoza\\ 
Calle Pedro Cerbuna, 12, 50009 Zaragoza, Spain}
\affiliation[f]{
Fermi National Accelerator Laboratory\\ 
Batavia, Illinois 60510, USA}
\affiliation[g]{
Instituto de Instrumentaci\'on para Imagen Molecular (I3M), Centro Mixto CSIC - Universitat Polit\`ecnica de Val\`encia\\
Camino de Vera s/n, 46022 Valencia, Spain}
\affiliation[h]{
LIBPhys, Physics Department, University of Coimbra\\
Rua Larga, 3004-516 Coimbra, Portugal}
\affiliation[i]{
Lawrence Berkeley National Laboratory (LBNL)\\
1 Cyclotron Road, Berkeley, California 94720, USA}
\affiliation[j]{
Instituto Gallego de F\'isica de Altas Energ\'ias, Univ.\ de Santiago de Compostela\\
Campus sur, R\'ua Xos\'e Mar\'ia Su\'arez N\'u\~nez, s/n, 15782 Santiago de Compostela, Spain}
\affiliation[k]
{Centro de Investigaci\'on en Ciencias B\'asicas y Aplicadas, Universidad Antonio Nari\~no\\ 
Sede Circunvalar, Carretera 3 Este No.\ 47 A-15, Bogot\'a, Colombia}
\affiliation[l]{
Department of Physics and Astronomy, Iowa State University\\
12 Physics Hall, Ames, Iowa 50011-3160, USA}
\affiliation[m]{
Department of Physics, University of Texas at Arlington\\
Arlington, Texas 76019, USA}
\affiliation[n]{
Departamento de F\'isica Te\'orica, Universidad Aut\'onoma de Madrid\\
Campus de Cantoblanco, 28049 Madrid, Spain}
\affiliation[o]{
Escola Polit\`ecnica Superior, Universitat de Girona\\
Av.~Montilivi, s/n, 17071 Girona, Spain}
\affiliation[p]{
Department of Physics and Astronomy, Texas A\&M University\\
College Station, Texas 77843-4242, USA}
\affiliation[q]{
Joint Institute for Nuclear Research (JINR)\\
Joliot-Curie 6, 141980 Dubna, Russia}
\emailAdd{ander.simon@ific.uv.es}

\abstract{The goal of the NEXT experiment is the observation of neutrinoless double beta decay in $^{136}$Xe using a gaseous xenon TPC with electroluminescent amplification and specialized photodetector arrays for calorimetry and tracking. The NEXT Collaboration is exploring a number of reconstruction algorithms to exploit the full potential of the detector. This paper describes one of them: the Maximum Likelihood Expectation Maximization (ML-EM) method, a generic iterative algorithm to find maximum-likelihood estimates of parameters that has been applied to solve many different types of complex inverse problems. In particular, we discuss a bi-dimensional version of the method in which the photosensor signals integrated over time are used to reconstruct a transverse projection of the event. First results show that, when applied to detector simulation data, the algorithm achieves nearly optimal energy resolution (better than 0.5\% FWHM at the $Q$ value of $^{136}$Xe) for events distributed over the full active volume of the TPC.}

\keywords{Double-beta decay detectors, Neutrinoless double-beta decay, TPC, High-pressure xenon
chambers, Image reconstruction, NEXT-100 experiment, NEW detector, ML-EM, Gaseous imaging and tracking detectors}

\collaboration{\includegraphics[height=9mm]{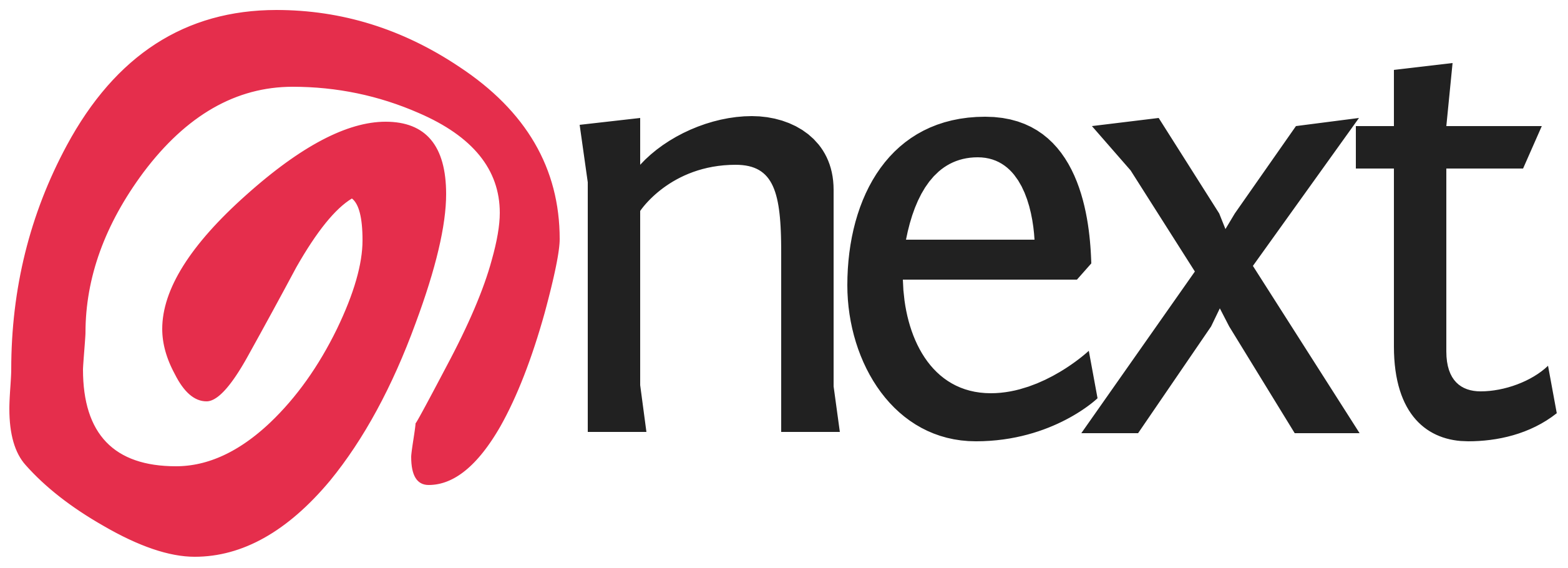}\\[6pt]
The NEXT Collaboration}

\begin{document}
\maketitle
\flushbottom

\section{The NEXT experiment}
\label{sec:next}

Neutrinoless double beta ($\beta\beta0\nu$) decay is a hypothetical process where two neutrons in a nucleus transform into two protons emitting only two electrons and no neutrinos. The detection of such a decay would demonstrate that neutrinos are Majorana particles and that total lepton number is not conserved \cite{GomezCadenas:2011it, Avignone:2007fu}. No experimental evidence of the decay has been found so far, with the most sensitive searches estimating the half-life of the decay, in the case of the $^{136}$Xe isotope, to be longer than $1.07\cdot 10^{26}$ years \cite{cite:Kamland}.

\begin{figure}[htbp]
	\centering
	\includegraphics[width=0.495\textwidth]{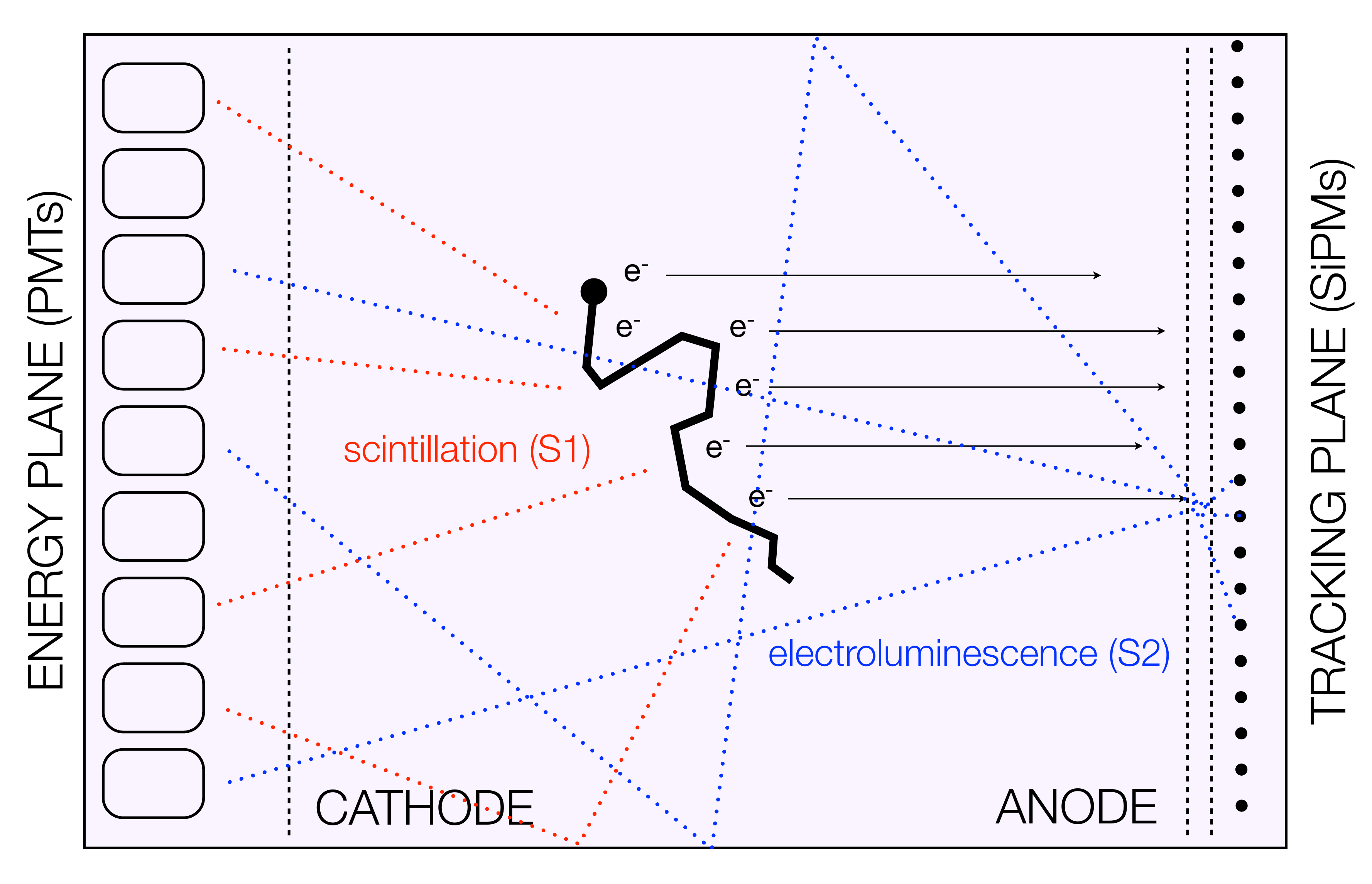}
	\caption{\label{fig:nextInteraction} Light production and detection scheme in NEXT}		
\end{figure}

The \textit{Neutrino Experiment with a Xenon TPC} (NEXT) \cite{cite:Martin-Albo:2016} will search for $\beta\beta0\nu$ decay using a high-pressure gaseous xenon time projection chamber. A schematic of the detection concept is shown in Figure \ref{fig:nextInteraction}. Charged particles passing through the active volume of the chamber ionize and excite the gas, causing a prompt scintillation signal that we call S1. The ionization signal is amplified, after drifting to the anode, using the electroluminescence of xenon, that is, the emission of proportional, secondary scintillation light after atomic excitation by a charge accelerated by a intense electric field. The difference in time between this second scintillation signal, called S2, and S1 gives the longitudinal position of the event. The electroluminescence region is located close to a tracking plane composed of SiPM detectors, while the so-called energy plane, which uses PMTs to detect the light, is on the opposite side of the chamber. This way, the light detected by the tracking plane is concentrated around the transverse position of the ionization signal, whereas the light arriving at the PMTs is diffuse and only weakly dependent on the transverse position of the signal.

The experimental signature of $\beta\beta0\nu$ decay consists in a fixed energy deposition equal to the energy difference between the decaying nucleus and its daughter. In addition, if produced in high pressure gas, the two emitted electrons deposit their energy at a known rate along erratic paths due to multiple scattering, until they become non-relativistic. At the end of the trajectory, a higher energy deposition occurs (Bragg peak). This behaviour is unique to double beta decay and can be used to discriminate signal (double electron events with two end-points of higher energy) from background (mainly single electron events with only one end-point of higher energy).

The NEXT detector concept provides excellent energy resolution, close to 0.5\% at the $Q$ value of $^{136}$Xe \cite{cite:DEMO,cite:DBDM}. The unique trace left in the detector by the decay makes track reconstruction an additional powerful handle to reject background \cite{Ferrario:2015kta}.

\begin{figure}[htbp]
	\centering 
	\includegraphics[height=0.4\textwidth]{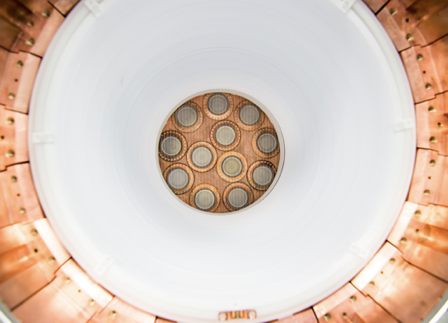}
	\includegraphics[height=0.4\textwidth]{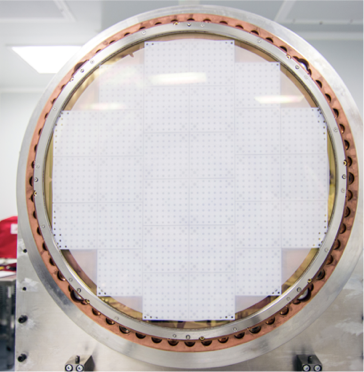}
	\caption{\label{fig:next} Energy (left) and tracking (right) plane of NEW.}
\end{figure}

The Collaboration is now starting the data taking of the NEXT-White (NEW) detector \cite{cite:NEW}. NEW is located underground at Laboratorio Subterr\' aneo de Canfranc (LSC) and its main objective is to corroborate the NEXT background model as well as to demonstrate the energy resolution for high energy events and evaluate the topological discrimination applicable due to track reconstruction. It will operate with 5 kg of depleted xenon at a pressure between 5 and 15 bar in a cylindrical vessel with 200 mm radius and 500 mm length. NEW is equipped with 12 PMTs and $\sim$2000 SiPMs (fig. \ref{fig:next}) and serves as the first phase of NEXT-100, which will be built by 2019. In the paper we will discuss and show how reconstructing events with ML-EM can enhance NEXT's tracking and energy resolution.

\section{Maximum Likelihood Expectation Maximization}
\label{sec:mlem}

Reconstruction in NEXT is a fairly complex problem since the information of two different planes have to be combined and there are many different effects that have an impact on the signal. In addition to that, the signal itself is rather complex by itself from a topological point of view. In this situation, methods capable of solving the inverse problem become of interest. In this scenario Maximum Likelihood Expectation Maximization arises as an ideal method for achieving a satisfactory event reconstruction because the distribution of statistical fluctuations are considered correctly within the method. This fact is of utmost importance when evaluating Poisson statistics which drive the photon counting processes.

\subsection{Maximum Likelihood Estimation}
\label{sec:mlem:ml}
Maximum Likelihood Estimation (MLE) \cite{cite:MLE} is a method suitable for solving a vast variety of inverse problems. As its name implies, the underlying idea behind it is to maximize the likelihood of a given statistical model (either analytic, simulated or derived from measurements) describing a forward problem, that is, the group of transformations that give a particular output from a determined input. 

Applying the method to a given data set corresponding to the output of the underlying statistical model, MLE provides estimates for the model's parameters. These parameters are the most likely solution, i.e.\, the most probable input that led to the observed measurements. In the case of a detection process, the data set provided to MLE would be the response of the detectors and MLE would estimate the source of that signal.

The log-likelihood function is preferred when using MLE since products are difficult to treat mathematically and because the logarithm is a monotone function. Therefore, the mathematical expression that defines MLE is:
\begin{equation}
\boldsymbol{x}_{\text{ML}} = \arg\max\log\mathcal{L}(\boldsymbol{x}|\boldsymbol{y}),
\label{eq:ml}
\end{equation}
where $\boldsymbol{x}_{\text{ML}}$ is the ML estimate, $\mathcal{L}$ is the likelihood function, $\boldsymbol{y}$ is the vector of observations and $\boldsymbol{x}$ is the vector of unknown parameters. Unfortunately, even with this consideration, analytical solutions are extremely rare, and maybe non-existent. As a consequence, a variety of numerical algorithms can be used to calculate MLE, being the Expectation Maximization method the most suitable.
\subsection{Expectation Maximization}
\label{sec:mlem:em}
Expectation Maximization (EM) \cite{cite:EM} is an iterative algorithm used for finding maximum likelihood estimates of the parameters of a given statistical model and a group of data. The algorithm is especially suited for problems with incomplete data or if the maximization of the likelihood function cannot be written explicitly. In this latter case, hidden variables can be introduced purely as a mathematical tool in order to make the estimation tractable. Given that, in the case of discrete variables, likelihood and probability can be related. This is done by adding the hidden variables in such a way that the knowledge of them considerably simplifies the maximization:

\begin{equation}
\mathcal{L}(\boldsymbol{x}|\boldsymbol{y}) = P(\boldsymbol{y}|\boldsymbol{x}) = \sum_{z}P(\boldsymbol{y},\boldsymbol{z}|\boldsymbol{x}),
\end{equation} 
where $\boldsymbol{z}$ is the vector of latent, or hidden, variables and $P(\boldsymbol{y},\boldsymbol{z}|\boldsymbol{x})$ is the marginal probability of the observed data.

If we know the value of the parameters $\boldsymbol{x}$, we can find the value of the latent variables $\boldsymbol{z}$ by maximizing the log-likelihood over all possible values of $\boldsymbol{z}$. Conversely, if we know the value of the latent variables $\boldsymbol{z}$, we can find an estimate of the parameters $\boldsymbol{x}$ fairly easily by simply grouping the observed data points according to the value of the associated latent variable. The algorithm works as follows:

\begin{enumerate}
\item Initialize the parameters $\boldsymbol{x}$ to some random values.
\item Expectation step (E-step): compute the best value for $\boldsymbol{z}$ given these parameter values.
\item Maximization step (M-step): Use the just-computed values of $\boldsymbol{z}$ to compute a better estimate for the parameters $\boldsymbol{x}$. 
\item Iterate steps 2 and 3 until convergence.
\end{enumerate}
Calculating the maximum likelihood estimator using this iterative method results in the Maximum Likelihood Expectation Maximization, ML-EM, whose mathematical expression and development depends on the likelihood model to be maximized \cite{cite:devMLEM}.

\section{Applying ML-EM in NEXT}
\label{sec:mlemNEXT}
Using ML-EM in NEXT requires the mathematical model that best describes the light production and detection processes inside the detector. In other words, development of the mathematical expression that derives from ML-EM algorithm and correct assessment of the observables and parameters of the method are needed.

\subsection{ML-EM expression in NEXT}
\label{sec:mlemNEXT:formula}
The ML-EM algorithm can be applied to several problems and, specifically, it has been broadly used in processes of Poisson nature. This is exactly the case of NEXT where scintillation and photon counting are the underlying processes. For a Poisson process, the log-likelihood function is: 
\begin{align*}
\log\mathcal{L}(\boldsymbol{x}|\boldsymbol{y}) &= \log\prod_{i}^{n}P(y_i|\boldsymbol{x}) \\
&= \log\prod_{i}^{n}\frac{e^{-\boldsymbol{Ax}|_i}(\boldsymbol{Ax}|_i)^{y_i}}{y_i!} \\
&= -\sum_{i}^{n}(\boldsymbol{Ax}|_i-y_i\log(\boldsymbol{Ax}|_i)+\log(y_i!),\numberthis\label{sec:mlem:poissonLikelihood}
\end{align*}
where $n$ is the dimension of the vector of observations $\boldsymbol{y}$ and $\boldsymbol{A}$ is the transformation matrix that relates the parameters $\boldsymbol{x}$ with the response $\boldsymbol{y}$, i.e.\ $\boldsymbol{y} = \boldsymbol{A}\cdot\boldsymbol{x}$, and $i$ indicates the ith vector component. After some rewriting and suppressing constant terms, equation \eqref{sec:mlem:poissonLikelihood} can be expressed as:

\begin{equation}
\log\mathcal{L}(\boldsymbol{x}|\boldsymbol{y}) = \sum_{i}^{n}(y_i\log(\boldsymbol{Ax}|_i)-\boldsymbol{Ax}|_i).
\label{sec:mlem:poissonFinal}
\end{equation}

In NEXT, the observed signal is the response of the photodetectors of both the energy and the tracking plane. Therefore, our output vector $\boldsymbol{y}$ will be the number of detected photons by each sensor $d$, $y_{d}$. This quantity of detections is directly proportional to the amount of light produced inside the detector which depends directly on the charge deposited in a given event, assuming, ideally, that there is no charge loss during the drift due to attachment. In other words, this deposited charge density is our input dataset, $\boldsymbol{x}$, and is the source of the registered response $\boldsymbol{y}$. This is true because only emission (Poisson process) and detection (Bernouilli process) are being considered, making the overall process driven by Poisson statistics.

In order to estimate this unknown density, the active volume of the TPC is voxelized (discretized) into small volume units, called voxels, in which the event charge is deposited. Then $\boldsymbol{x}$ is a vector whose components are proportional to the number of ionization pairs $x_{v}$ produced at each voxel $v$. 

Given the Poisson nature of the detection process, all steps for derivation of the ML-EM in nuclear medicine apply directly to the NEXT case. As a consequence, expression \eqref{sec:mlem:poissonFinal} can be solved using the same Expectation Maximization algorithm used in medical imaging \cite{cite:MLEM}. This results, writing the sums and probabilities explicitly, in:
\begin{equation}
x_m(v) = \frac{x_{m-1}(v)}{\sum_d P(v,d)}\sum_d \frac{y(d)P(v,d)}{\sum_{v'} x_{m-1}(v')P(v',d)}
\label{eq:mlemNEXT}
\end{equation}
where $x(v)$ is the charge deposited in the voxel $v$, $P(v,d)$ is the probability of detection by the detector $d$ when having a photon emitted due to the charge deposited in voxel $v$ and $y(d)$ is the number of photoelectrons produced in the detector $d$. This is the most basic and applicable expression of the ML-EM algorithm to NEXT.

The denominator inside the sum over the detectors in expression \eqref{eq:mlemNEXT} is an estimation of what the sensor should be measuring with the current estimates of the parameters, that is, with the current charge distribution estimated. This is known as the forward projection and noise can be easily accomodated into it by simply adding a noise term \footnote{Analogous to the implementation in \cite{cite:MLEMnoise}}, leaving \eqref{eq:mlemNEXT} as:

\begin{equation}
x_m(v) = \frac{x_{m-1}(v)}{\sum_d P(v,d)}\sum_d \frac{y(d)P(v,d)}{\big(\sum_{v'} x_{m-1}(v')P(v',d)\big) + c(d)}
 \label{eq:mlemNEXTNoise}
\end{equation}
where $c(d)$ is the mean noise of the detector $d$. 

\subsection{Implementation of ML-EM}
\label{sec:mlemNEXT:implementation}
The variables and parameters from expression \eqref{eq:mlemNEXTNoise} have to be clearly established and defined in order to apply the method. In addition to that, for a computationally efficient implementation of the algorithm, several simplifications can be made to the problem to improve the overall speed and performance of the method without reducing its precision.

\subsubsection{Probability model}
\label{sec:mlemNEXT:implementation:probability}
To apply the ML-EM expression it is necessary to know the probability model $P(v,d)$. This can be computed by simulation because an analytic description of all physical effects is very challenging. For this the NEXT detector simulation is used. A high number of electrons, to reduce statistical noise, are simulated inside the detector volume at an exact location $q(x,y,z)$, thus producing light and generating a response in the sensors. In this case, the number of photons detected divided by the number of photons produced is an estimate for the probability of light detection of sensor $d$ from an electron deposited in point $q$. This is done for each point $q$ inside an imaginary grid which divides uniformly the detector volume. As a result, a matrix of probabilities $P(q,d)$ is obtained. If the voxels are small enough, then the probability $P(v,d)$ from a voxel can be approximated to that of the point $q$ in the center of the voxel, $P(v,d) \approx P(q,d)$.

The situation above explained would be the ideal approach. This matrix, however, is too big to handle. In order to reduce its size and computation time, it is only produced for electrons at the EL region, that is, the probability matrix only accounts for points $q(x,y,z_{EL})$. This is valid since all the ionization electrons produced inside the chamber drift towards the EL region and light is produced inside this region. The approach neglects the drift effects of charge distribution but those are outside the scope of this paper and can be added analytically at a later stage.

Even with this simplification, the probability matrix is rather large. An imaginary transverse section of the TPC is divided in a 1 per 1 mm grid. The response of the light sensors is simulated for light produced at each of those cells. The number of cells is more than 150 thousand in the case of NEW, each one with a number of probabilities associated equal to the number of sensors that detect light from that point. This situation still leads to large probability matrix that would become even larger when moving to larger detectors.

To avoid this issue, a parametrization of the forward problem can be derived from the simulations. It allows to predict the probability matrix with a much higher granularity using a probability function that depends on the distance between the point and the sensor. 

\begin{figure}[htbp]
	\centering
	\includegraphics[width=.495\textwidth]{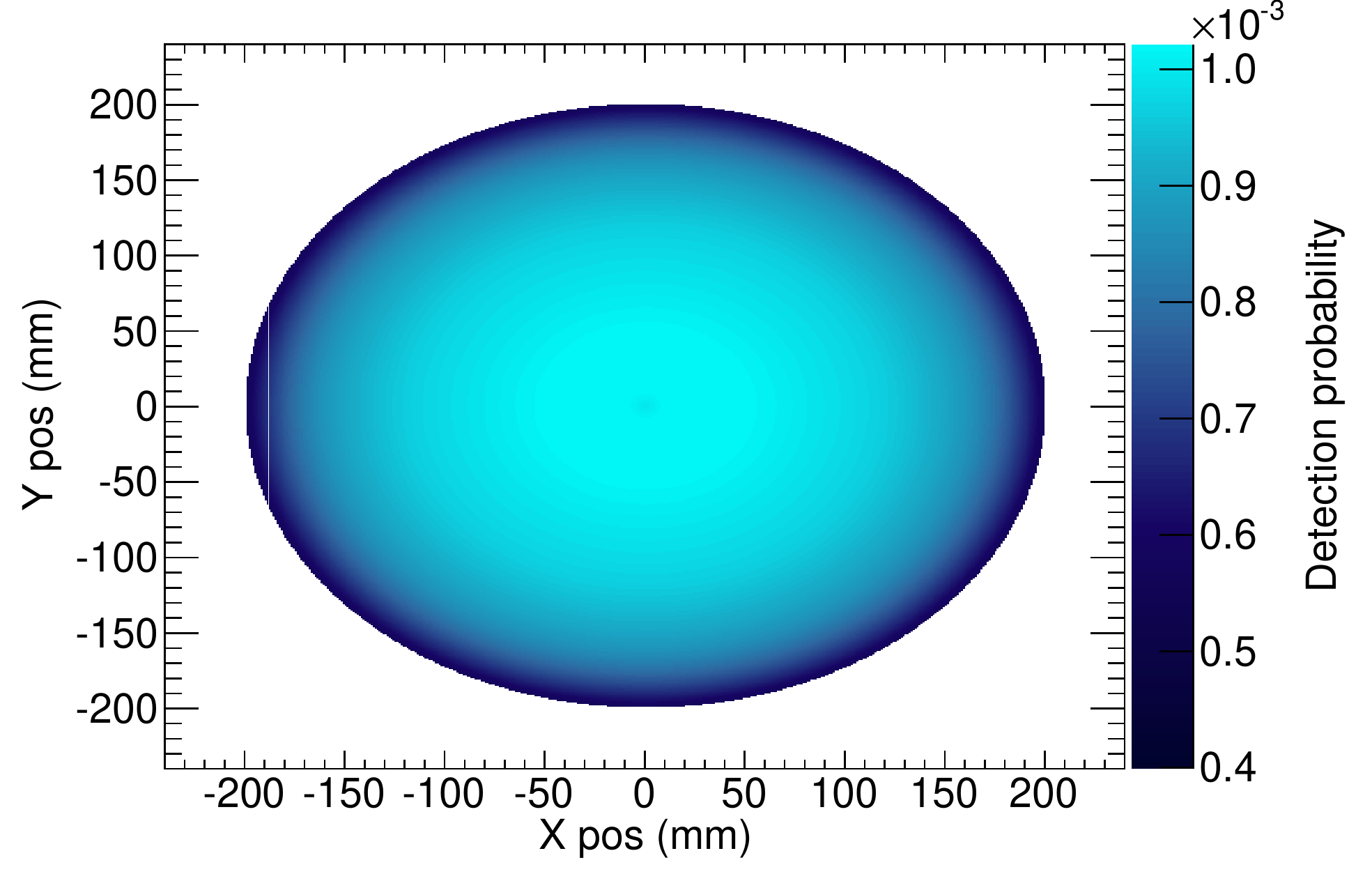}
	\includegraphics[width=.495\textwidth]{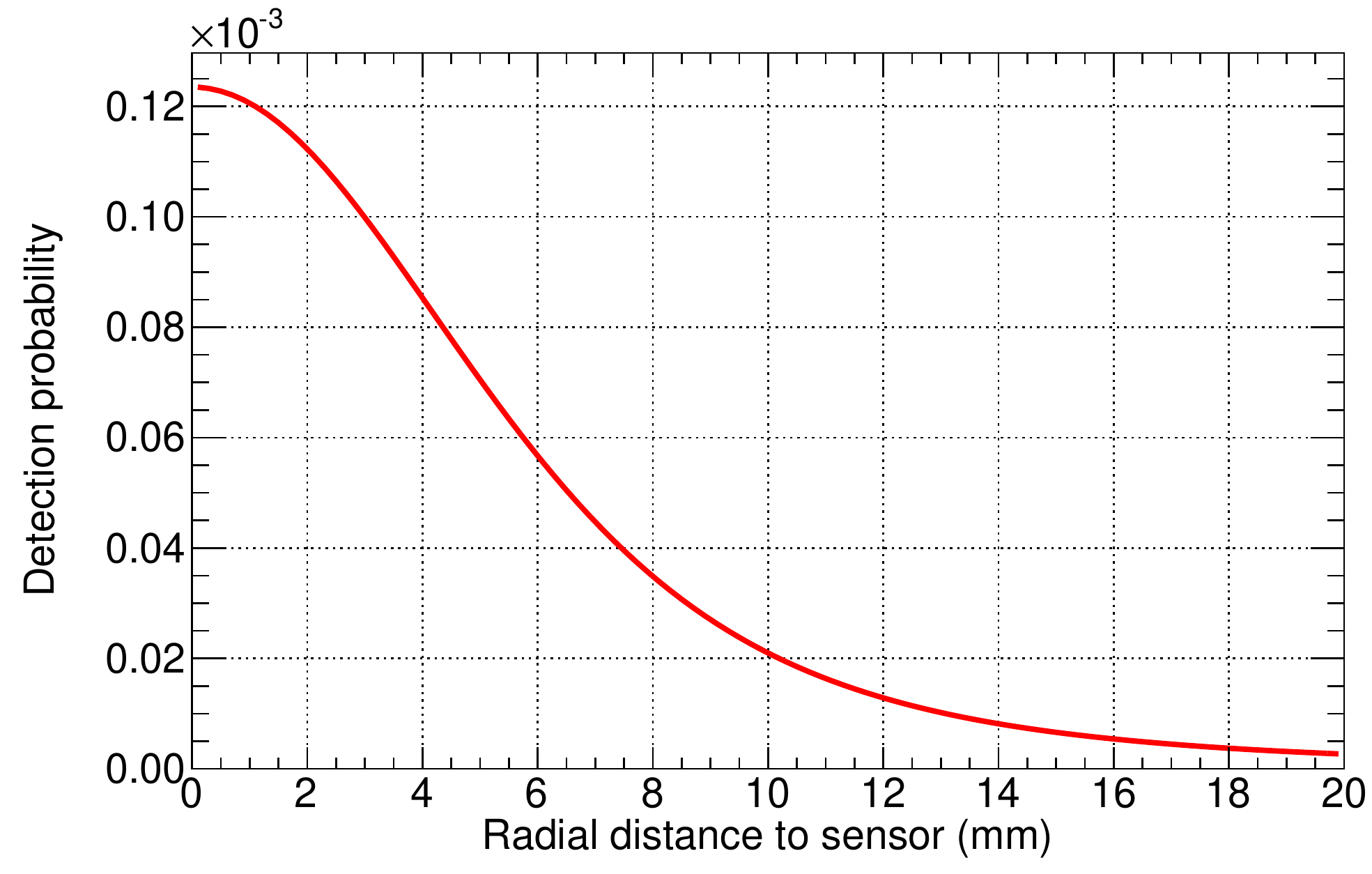}
	\caption{\label{fig:prob} Light detection probability function of the PMTs along the XY position (left) and of the SiPMs along the radial distance to the sensors (right).}
\end{figure}

The parametrization of the PMTs is done individually and a function is obtained for each one of the twelve channels. This is done by profiling the detection probabiliy along the radial position of the points relative to the longitudinal axis of the chamber. Then, a profile of this profile is done alongside the phi position of the event for several radial positions. With this, an XY distribution of probabilities is obtained (fig.\ \ref{fig:prob}, left).

In the case of the SiPMs, an individual function is not optimal since it would mean having almost 2000 functions for NEW and 4000 functions for NEXT-100, the next phase of the NEXT experiment. A different approach is followed: instead of using the radial distance from the point to the XY center of the chamber, the distance of the point to the sensor is used. Then, a profile of the probabilities relative to the distance sensor-point can be done. The profile is only done for distances up to 20 mm, further from that the probability of detection is considered negligible since light is concentrated only within the sensors positioned in front of the event (probability 20 mm away is approximately 2\% of the probability in front of the sensor). A spline fit of this profile serves as the probability function for the SiPMs (fig.\ \ref{fig:prob}, right). 

\subsubsection{Reconstruction mode}
\label{sec:mlemNEXT:implementation:modes}
Using probability functions that do not account for drift effects constrains the tridimensional viability of the method. However, even with this limitations, the algorithm can be applied and used for reconstruction. With the current probability functions a two-dimensional reconstruction mode has been developed.

In this mode, the signal of the sensors is fully integrated over time losing information of the evolution of the event alongside the longitudinal axis. As a consequence, we can apply \eqref{eq:mlemNEXTNoise} to the integrated charge $n(d)$ and obtain the charge distribution in the EL plane. This simplification of the problem reduces its computational charge. Consequently, instead of voxelizing the full volume of the chamber, we pixelize (that is, we discretize in area units) the EL region.

As a result, the output of the method is a collection of pixels that reflect the XY position of the ionization electrons when they arrived to the EL area and the number of them that have ended up drifting to that pixel, thus it is an image of the original XY projection of the track convoluted with drift effects. In this case we have no z-dimension information besides the mean position given by the difference in time between S2 and S1. In addition to this 2D track, the sum of all the pixels' energy can be used as an estimate of the energy of the event.

\subsubsection{Convergence of the method}
\label{sec:mlemNEXT:implementation:convergence}
Convergence of the expression \eqref{eq:mlemNEXT} has already been proven \cite{cite:MLEM}. Although it may seem convenient to iterate as much as possible, ML-EM suffers from a noise increment when over-iterating \cite{cite:EMnoise1, cite:EMnoise2}; therefore a stopping criteria has to be defined.

\begin{figure}[htbp]
	\centering
	\includegraphics[width=0.495\textwidth]{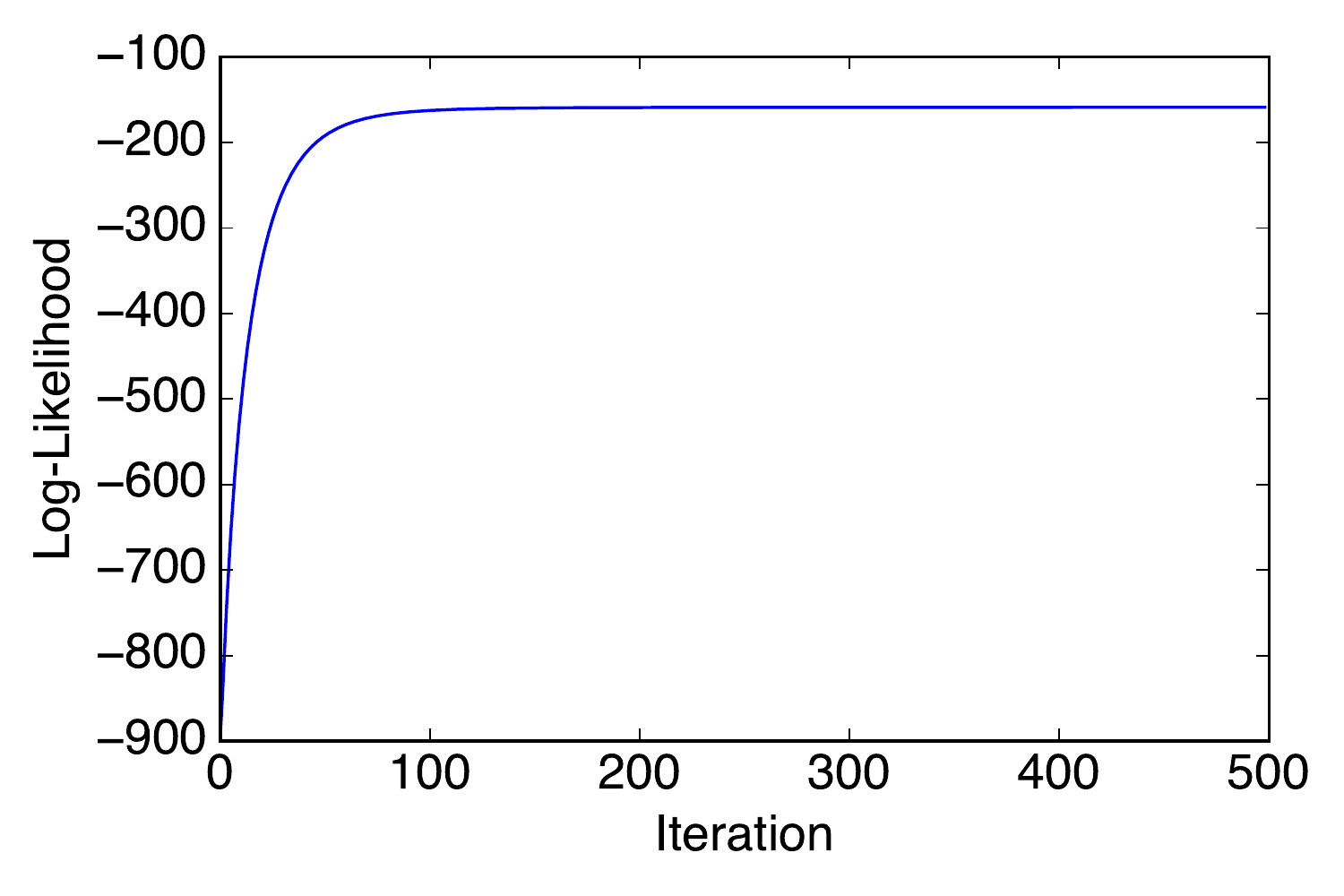}
	\includegraphics[width=0.495\textwidth]{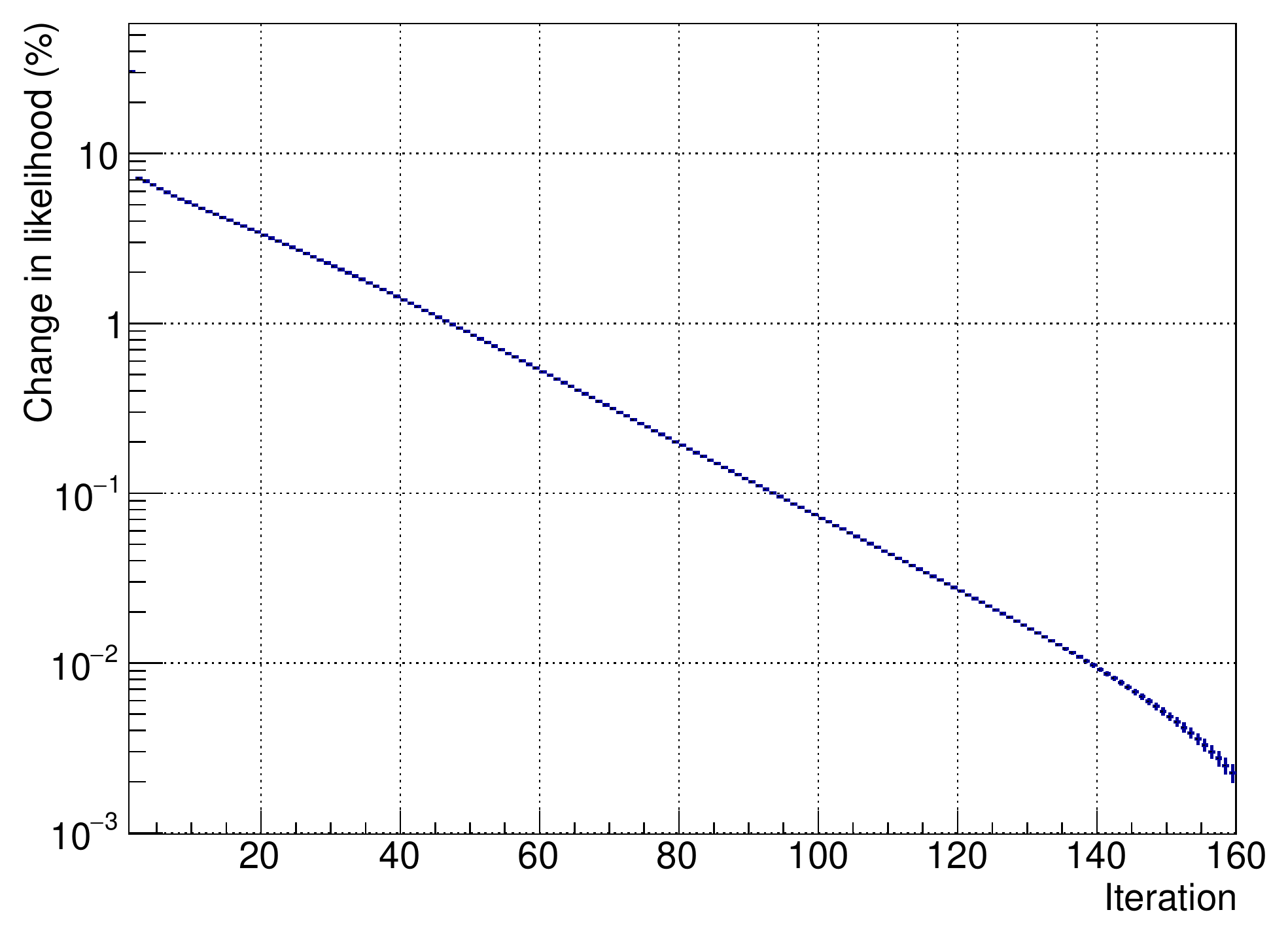}
	\caption{\label{fig:likelihood} Evolution of the likelihood for a event (left) and mean percentual change of the likelihood along the number of iterations (right).}
\end{figure}

The convergence of the likelihood can serve as the stopping criteria of the algorithm. Likelihood can be calculated at each iteration comparing the probability of having the detected sensor response with respect to a Poisson distribution with mean equal to the forward projection given by the method (fig.\ \ref{fig:likelihood}, left). Studying the percentage change in likelihood can provide then a valid threshold to end the iterative process. By reconstruction of 10 000 simulated events (fig.\ \ref{fig:likelihood}, right), we found that a cut below 0.1\% is reasonable for stopping the iterations. This change usually is reached slightly before 100 iterations are done.  

\subsubsection{Method initialization}
\label{sec:mlemNEXT:implementation:init}

A seed is needed to initialize any iterative method such as ML-EM. In order to avoid any prior bias, the initial charge of pixels is set as a uniform distribution. The initial value of this distribution, as long as it is uniform, can be arbitrary. Our choice has been to use the total number of photons detected by PMTs divided by the total number of pixels that will be iterated through.

Silicon photomultipliers probability function implies that SiPM only detect non-negligible light if it is produced near them. Keeping this in mind, only sensors with non-zero detection probability are considered to reduce computational charge. As a consequence, for the starting number of pixels, a region of interest is selected instead of pixelizing all the space. To do this, a rectangular area that covers the SiPMs that have detected light is selected. The sides of this rectangle are enlarged up to 20 mm to consider the pixels where, according to the SiPM probability function (fig.\ \ref{fig:prob}, right), produced light could have been detected if produced. This latest consideration avoids any possible prior bias to the seed. 

Finally, the response vectors are defined. All PMTs are taken into account since all of them detect light. That is not the case with SiPMs because not all of them detect light nor have detection probability within the region of interested chosen.  Then, iterating over all of SiPMs is, again, ineffective and only those with non-negligible probability detection are considered.


\subsubsection{Noise implementation}
\label{sec:mlemNEXT:implementation:noise} 

\begin{figure}[htbp]
	\centering 
	\includegraphics[width=0.495\textwidth]{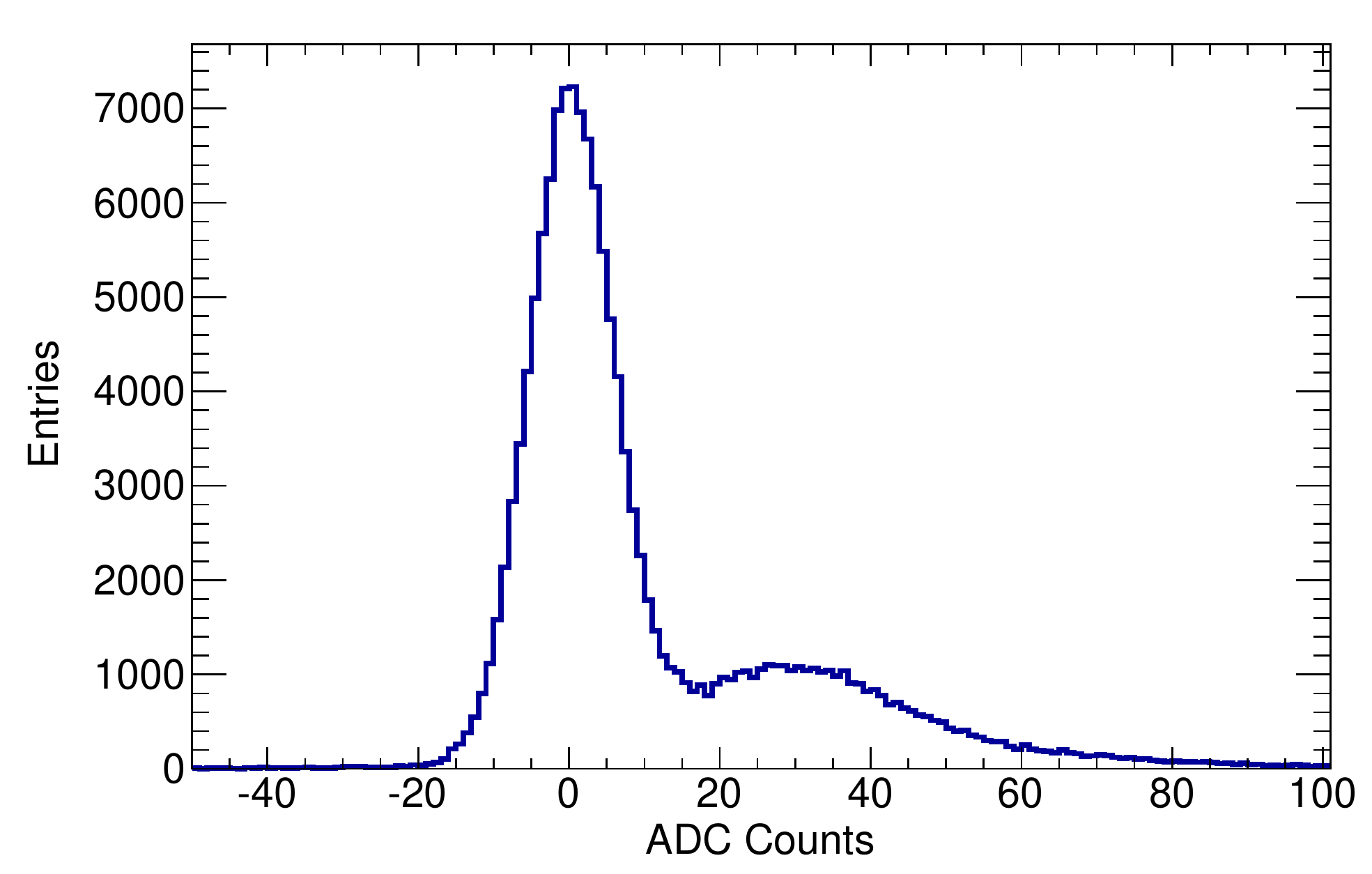}
	\includegraphics[width=0.495\textwidth]{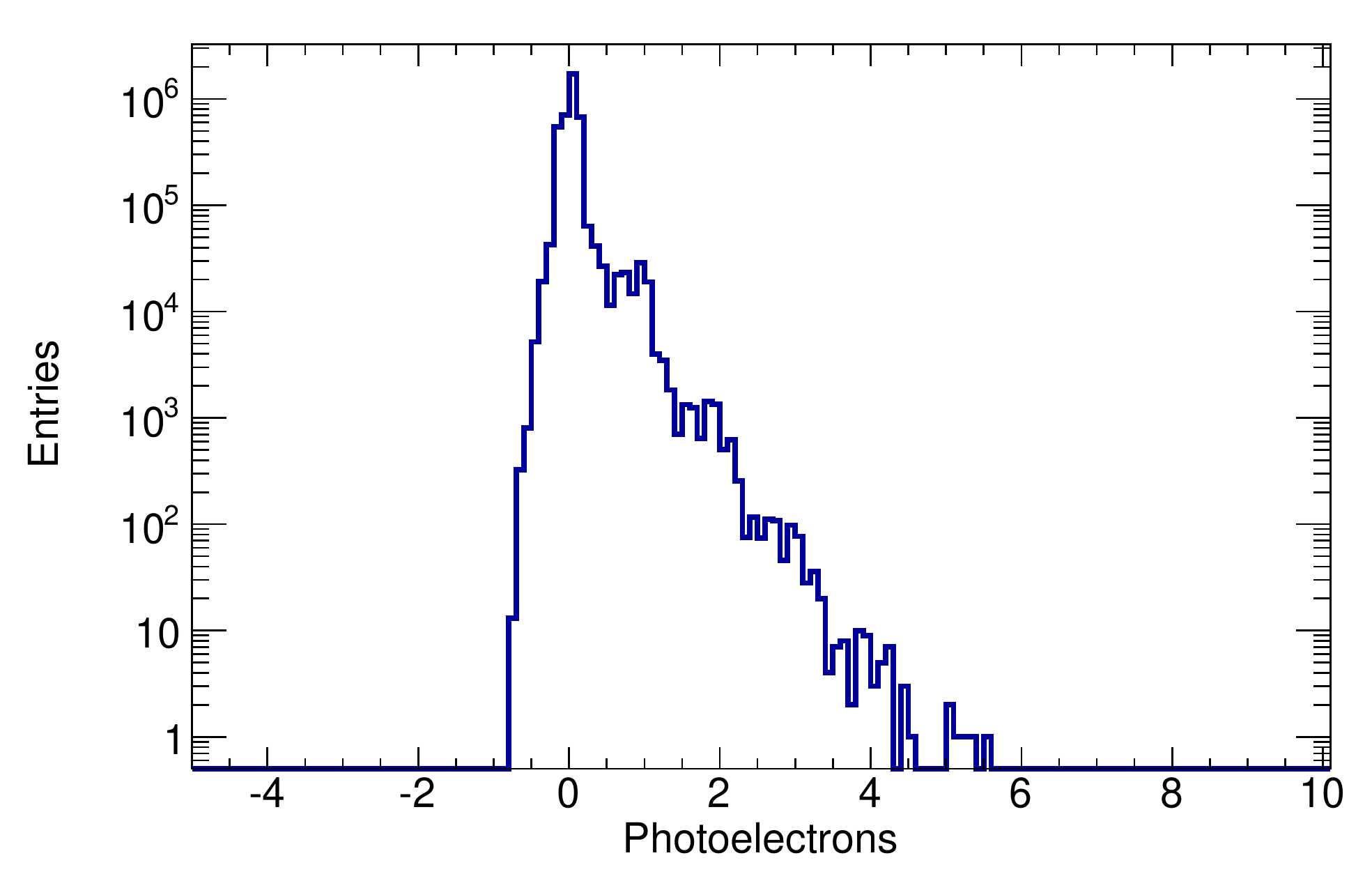}
	\caption{\label{fig:noise} Single photoelectron spectrum of a PMT (left) and dark noise spectrum of a SiPM (right). }
\end{figure}

The mathematical expression of ML-EM in NEXT \eqref{eq:mlemNEXTNoise} includes the term $c(d)$ that accounts for sensor noise. This term translates into the mean noise of each sensor over the period of signal integration. In the case of the PMTs the noise is the baseline variation of the sensor. This variation, however, follows a gaussian distribution centered at zero, as shown on fig.\ (\ref{fig:noise}, left), with a RMS of 0.04 photoelectrons per 100 ns. Therefore, the mean noise of the PMTs is considered to be zero when applying ML-EM.

For the SiPMs, however, in addition to the noise from electronic sources and baseline variation, centered at zero, there is a strong thermal noise component, also known as dark noise. Then the noise term of the formula has to include both terms. This can be done by using the mean value of the dark noise spectrum of the sensor (fig.\ \ref{fig:noise}, right). The noise values shown indicate the rate per microsecond. To transform them into the total mean noise, taken as $c(d)$ in the ML-EM formula, they have to be multiplied by the full time interval of the event. 

\section{Simulated data generation}
\label{sec:sim}
The data used in the paper was produced using the NEW detector simulation of NEXT, based on Geant4 \cite{cite:GEANT}. It was generated using a fast simulation mode in which only charge depositions of the event are simulated. The effects of the charge drift can easily be included analytically later making the simulation process much faster. The light detected by each sensor is also included afterwards using the light detection probability parametrization. This means the simulated data and ML-EM use the same probability model except for the non-inclusion, at the moment, of charge drift effects in ML-EM. Therefore, the reconstruction obtained through this algorithm will be smeared by these effects.

Samples of $^{83}$Kr, $^{22}$Na, $^{208}$Tl and $^{136}$Xe $\beta\beta 0 \nu$ events have been produced. The first three correspond to the sources planned to be used for calibration in NEW while the latest corresponds to the signal events. The pressure of the xenon in the production is set to 15 bar. For this pressure a transverse diffusion of 1 mm/$\sqrt{\text{cm}}$ and a longitudinal diffusion of 0.3 mm/$\sqrt{\text{cm}}$ have been considered following previous measurements at 10 bar made by the Collaboration \cite{cite:DEMO}. The potential difference in the electroluminescence region is considered to be 13.73 kV which entails a E/p (coefficient between electric field applied and gas pressure) in the electroluminescence region of $\sim$1.83 kV/(cm$\cdot$bar). This E/p corresponds to a gain of 1050 photons per ionization electron \cite{cite:ELGain}. 

The charge resolution of the PMTs has been considered to be 0.6. Noise in the sensors is included in the signal generation by adding a random value to each time bin of the waveforms. For PMTs, noise is considered to be a random value following a gaussian distribution centered at zero with RMS equal to the actual baseline noise of each channel. In the case of SiPMs, to consider electronic and thermal noise, the noise is included by adding a random number following the dark noise spectrum distribution (fig.\ \ref{fig:noise}, right) measured for each sensor.

\section{Method performance studies}
\label{sec:perform}

As explained, the main goal of applying ML-EM in NEXT is to achieve good energy resolution and track reconstruction at the same time. With this in mind, evaluation of the algorithm parameters must be done in order to optimize the performance of those traits.

The most effective way of doing this evaluation is using the simplest possible case, i.e.\ point-like depositions. Broadly used for calibration purpouses \cite{Kastens:2009rt}, $^{83}$Kr low deposition energy (41.5 keV) translates into a point-like deposition making it ideal for the studies. 

Energy resolution itself is a good indicator of the performance and can be directly evaluated. For track reconstruction, the rms of the difference between the mean position of the reconstructed pixels weighted by their charge and the true position of the event is evaluated for our ML-EM implementation. The following studies have all been done using 100 000 events and the errors assigned are derived from the fits and weighted with the reduced $\chi^2$. 

\subsection{Optimal number of iterations}
\label{sec:perform:iter}

Beyond taking into account the estimator of the model explained in section (\ref{sec:mlemNEXT:implementation:convergence}), the maximum number of iterations applied must be chosen in a way where the performance approaches optimal values within a reasonable computational time.

\begin{figure}[htbp]
	\centering 
	\includegraphics[width=0.495\textwidth]{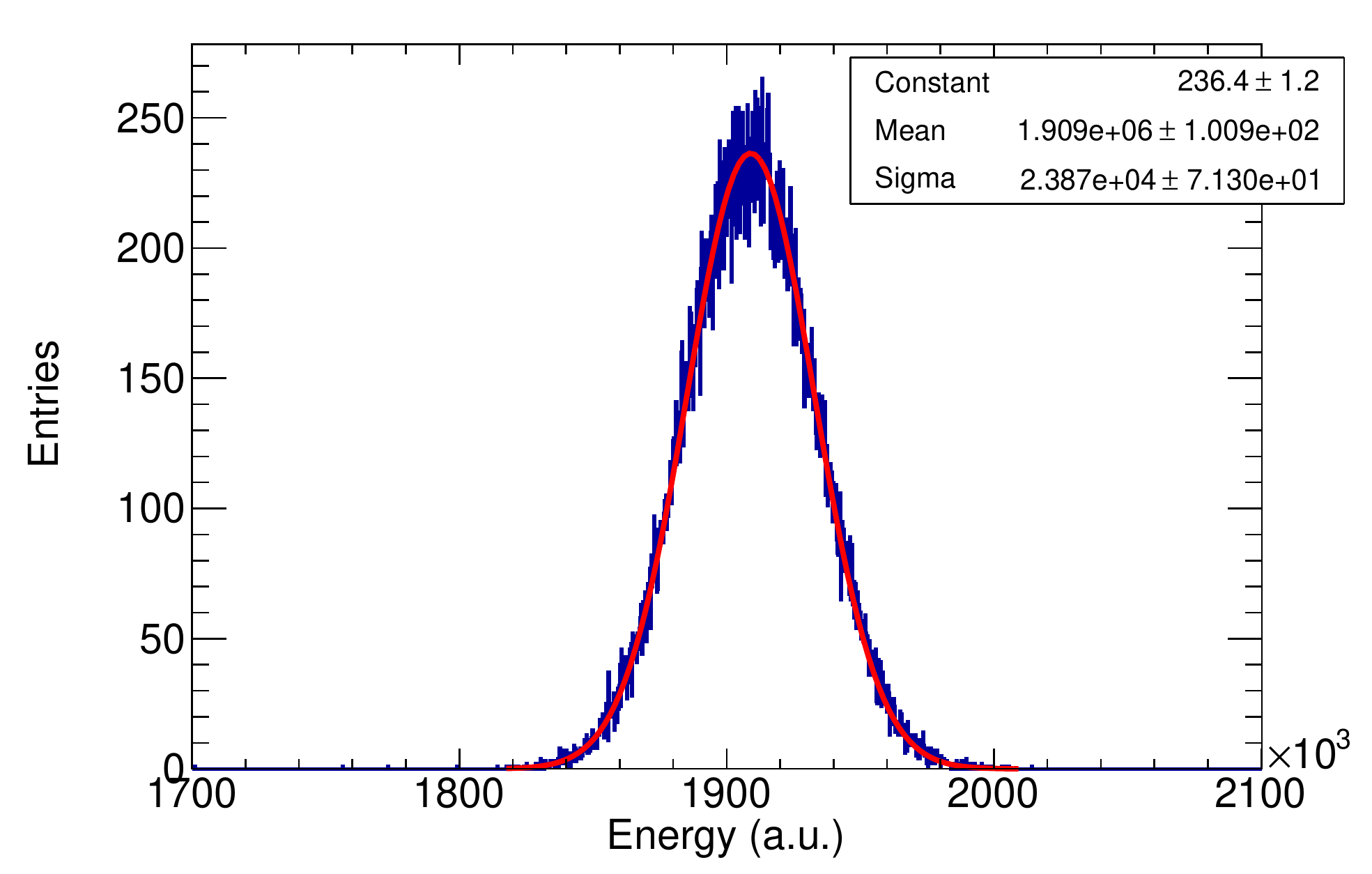}  
	\includegraphics[width=0.495\textwidth]{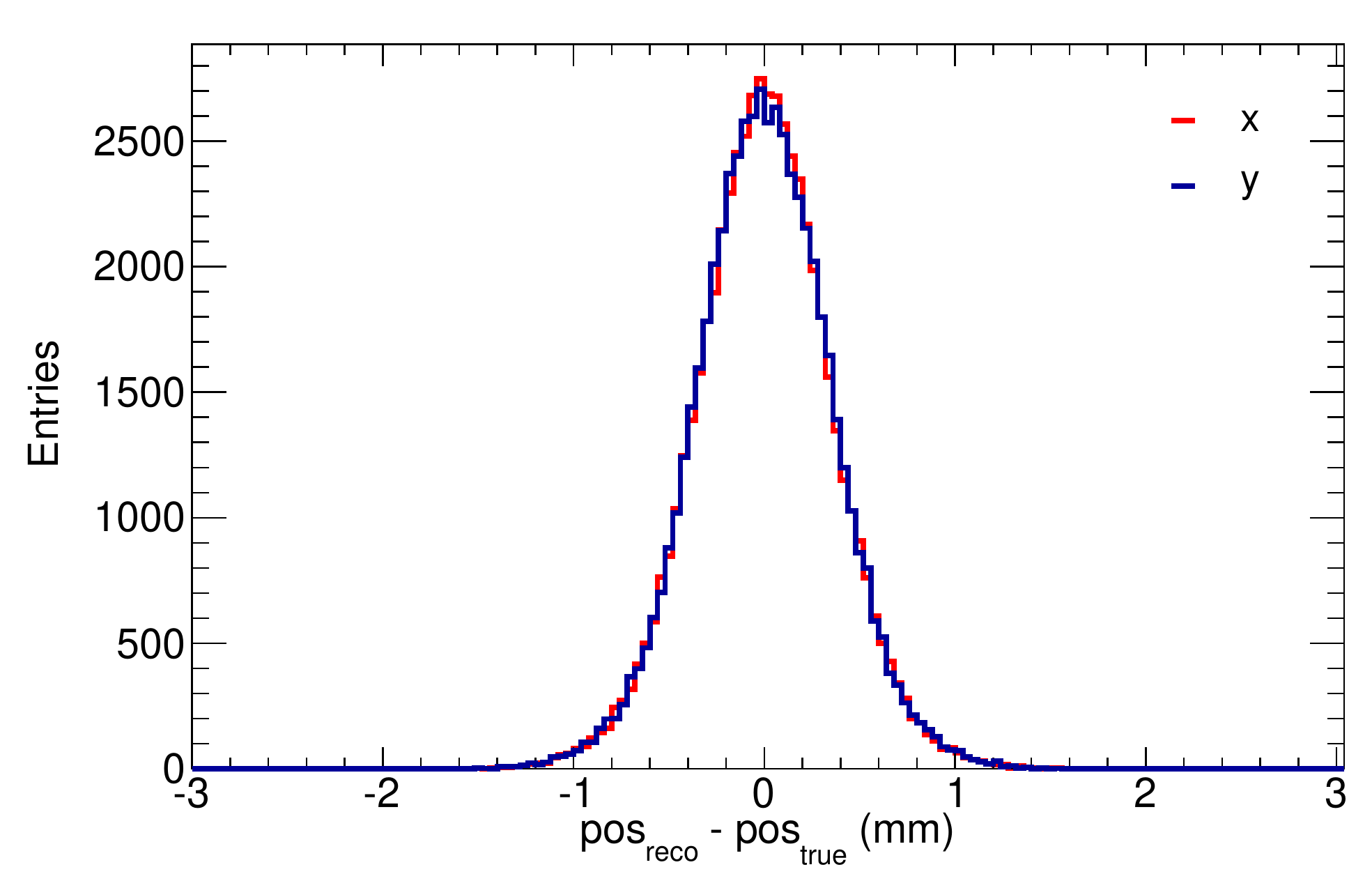}
	\caption{\label{fig:Iter100} Reconstructed energy of $^{83}$Kr (left) and distribution of the position difference for $x$ and $y$ axis between reconstruction and true position (right) after 100 iterations.}
\end{figure}

\begin{figure}[htbp]
	\centering 
	\includegraphics[width=0.495\textwidth]{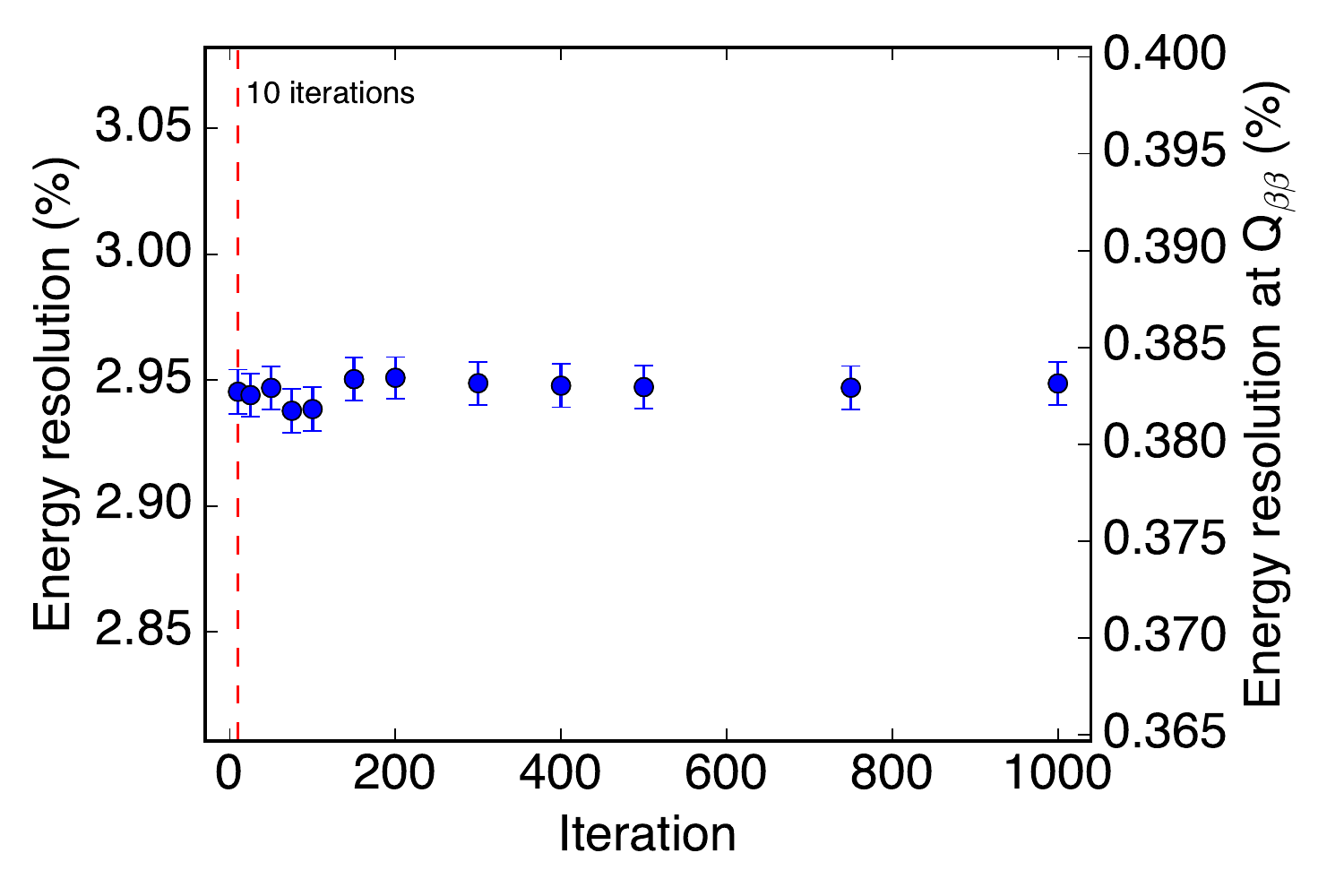}
	\includegraphics[width=0.495\textwidth]{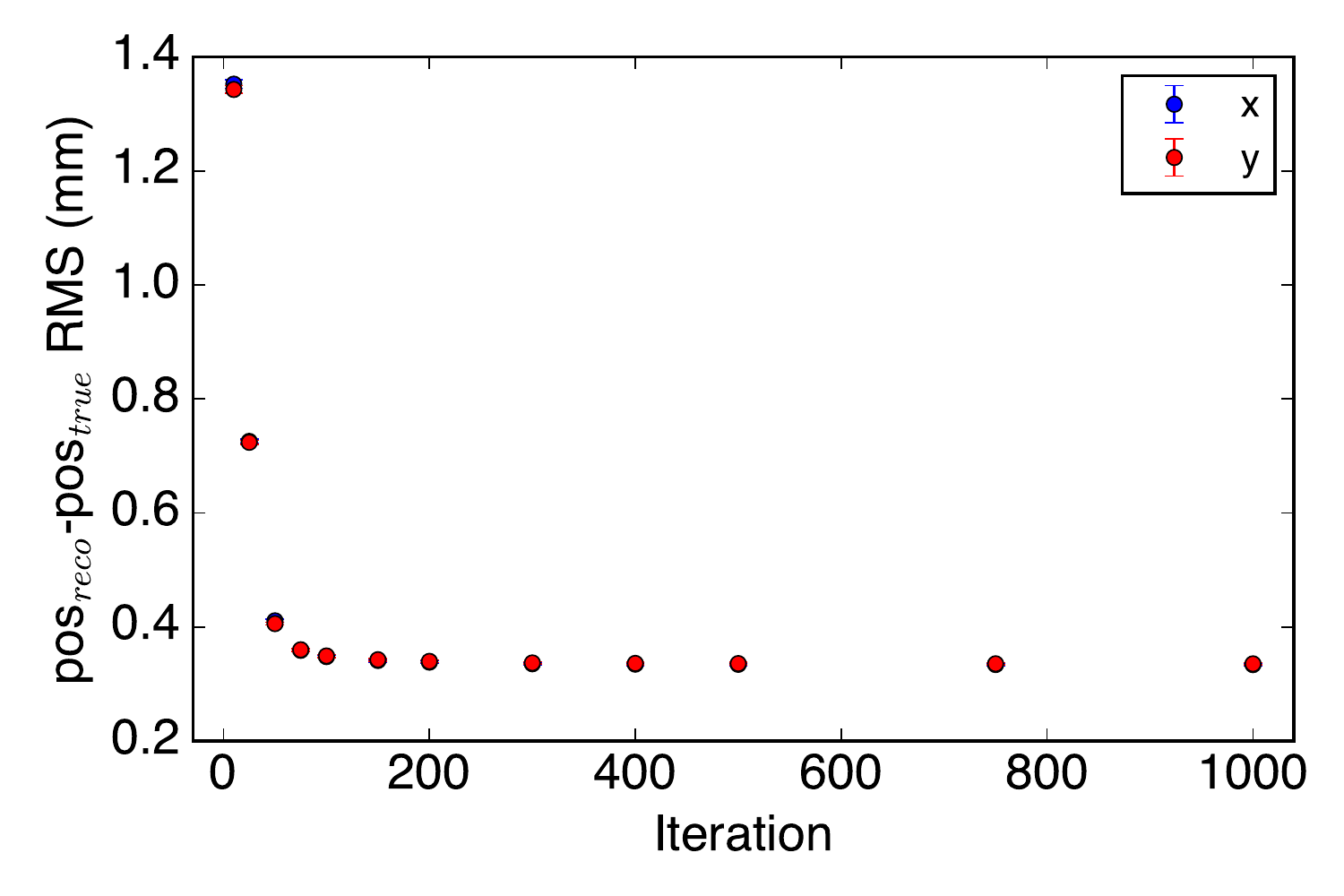}
	\caption{\label{fig:Iter} Dependence on the number of iterations of the energy resolution (left) and of the RMS of the XY mean position difference between reconstructed and true information (right) for $^{83}$Kr events.}
\end{figure}

Applying the method for several number of iterations using a pixel size of 10$\times$10 mm to events fully contained within 150 mm of the longitudinal axis of the detector shows that energy resolution reaches a mean value of $\sim$2.94 \% FWHM (fig.\ \ref{fig:Iter100}, left) at the krypton energy with only 10 iterations (fig.\ \ref{fig:Iter}, left).
The resolution obtained through ML-EM extrapolates to $\sim$0.38 \% at $Q_{\beta\beta}$, better than the Collaboration's optimistic goal. Increasing the number of iterations does not improve this number, therefore iterating more than this will only add unprofitable computation time. 

Difference between reconstructed mean position and mean true position also shows that around 100 iterations are enough to get satisfactory values (fig.\ \ref{fig:Iter}, right). At this number of iterations the distribution is peaked at almost zero with, after applying a gaussian fit, a sigma of around 0.4 mm in both dimensions (fig.\ \ref{fig:Iter100}, right). This implies that all the events are reconstructed within a radial distance of $\sim$ 1.7 mm from the real position with 3$\sigma$.

While energy resolution shows that only ten iterations are enough to get an stable and satisfactory performance, the position difference evaluation indicates that a cut at around 100 iterations is desirable.  This number is compatible with the likelihood (fig.\ \ref{fig:likelihood}, right) evolution alongside the number of iterations where the difference between iterations is below 0.1\%.


In addition to this, computation time has also to be taken into account. A fit to the computing time for several number of iterations indicates that each iteration of the ML-EM algorithm needs approximately 4.9 ms to be computed. 500 ms for 100 iterations are a reasonable time to make a fast event reconstruction and perform a quick energy cut in order to do a detailed study.


\subsection{Pixel size}
\label{sec:perform:voxel}

An adequate pixel size selection is also needed for optimal performance. A large pixel may enclose some of the geometrical effects that aims to be corrected, therefore worsening the capability of correction at such level. Smaller pixel size is, then, the desired objective. However, like with the case of iterations, this increases highly the computational time. Therefore, pixel size has to be evaluated and optimized. This evaluation is done applying 100 iterations and considering square pixels with sizes between 1 mm to 20 mm in 1 mm steps.

\begin{figure}[htbp]
	\centering 
	\includegraphics[width=0.495\textwidth]{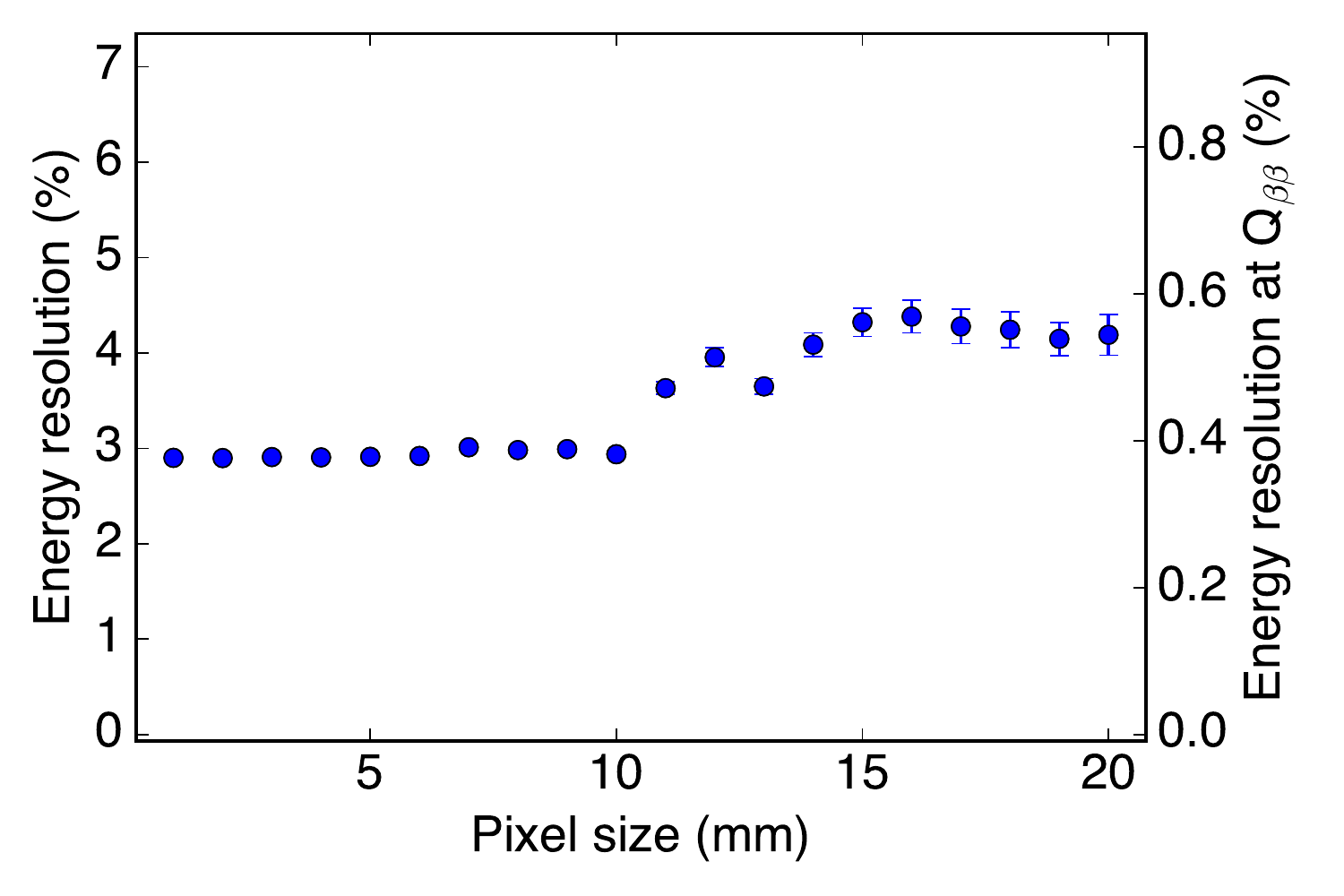}
	\includegraphics[width=0.495\textwidth]{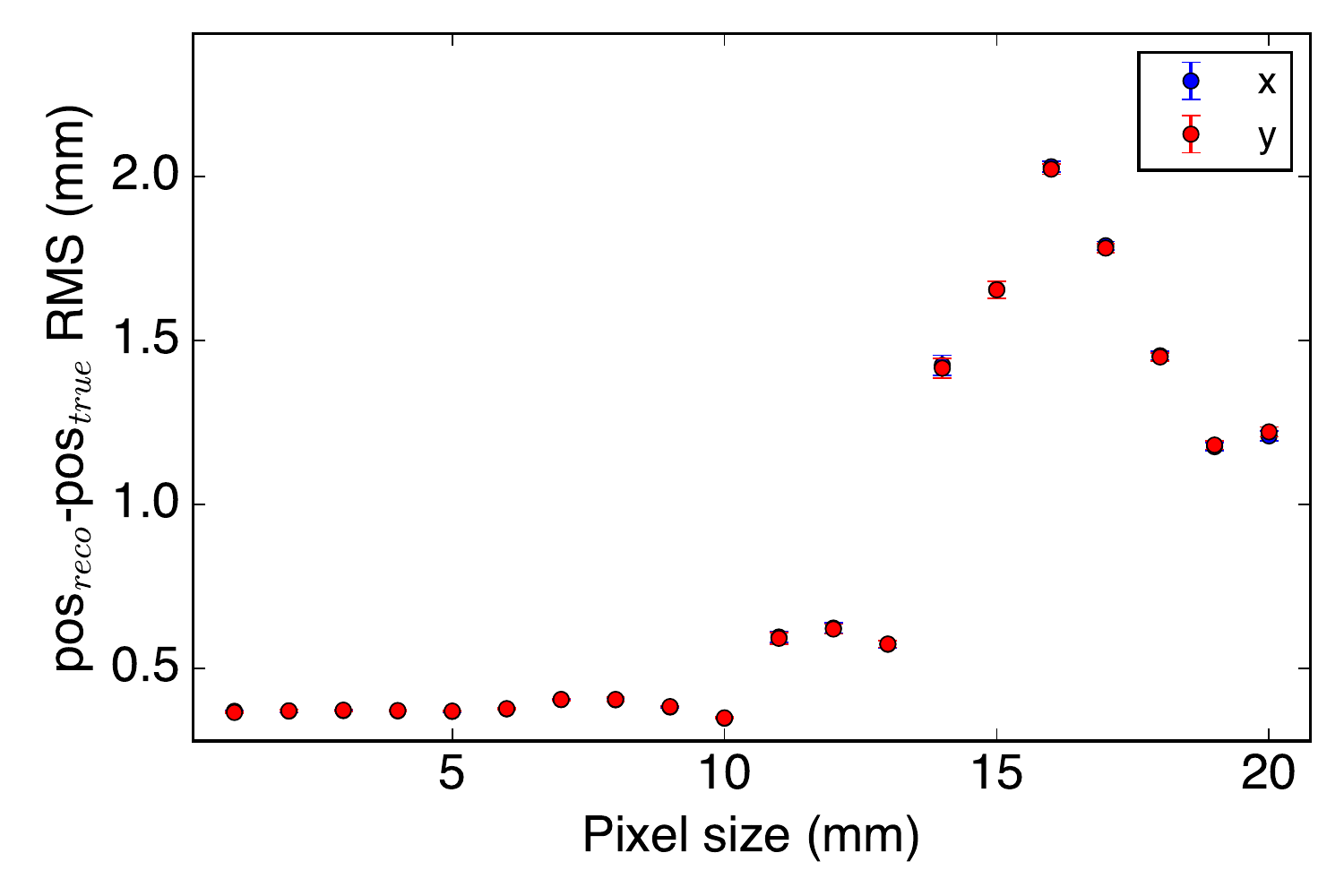}
	\caption{\label{fig:Size} Pixel size dependence of the energy resolution (left) and of the rms of the XY mean position difference between reconstructed and true information (right) for $^{83}$Kr.}
\end{figure}

Study of the change of energy resolution depending on the pixel size (fig.\ \ref{fig:Size}, left) shows that, up to 10 mm resolution, energy resolution remains stable. The same behaviour can be observed when looking at the difference between reconstructed and true position (fig.\ \ref{fig:Size}, right). However, for bigger sizes the method starts to notably worsen and to approximate the probability associated to the pixel to that of the center of the pixel is no longer valid. Therefore, any pixel size up to 10 mm is a valid choice without degrading the performance of the method.

\begin{figure}[htbp]
		\centering
		\includegraphics[width=0.6\textwidth]{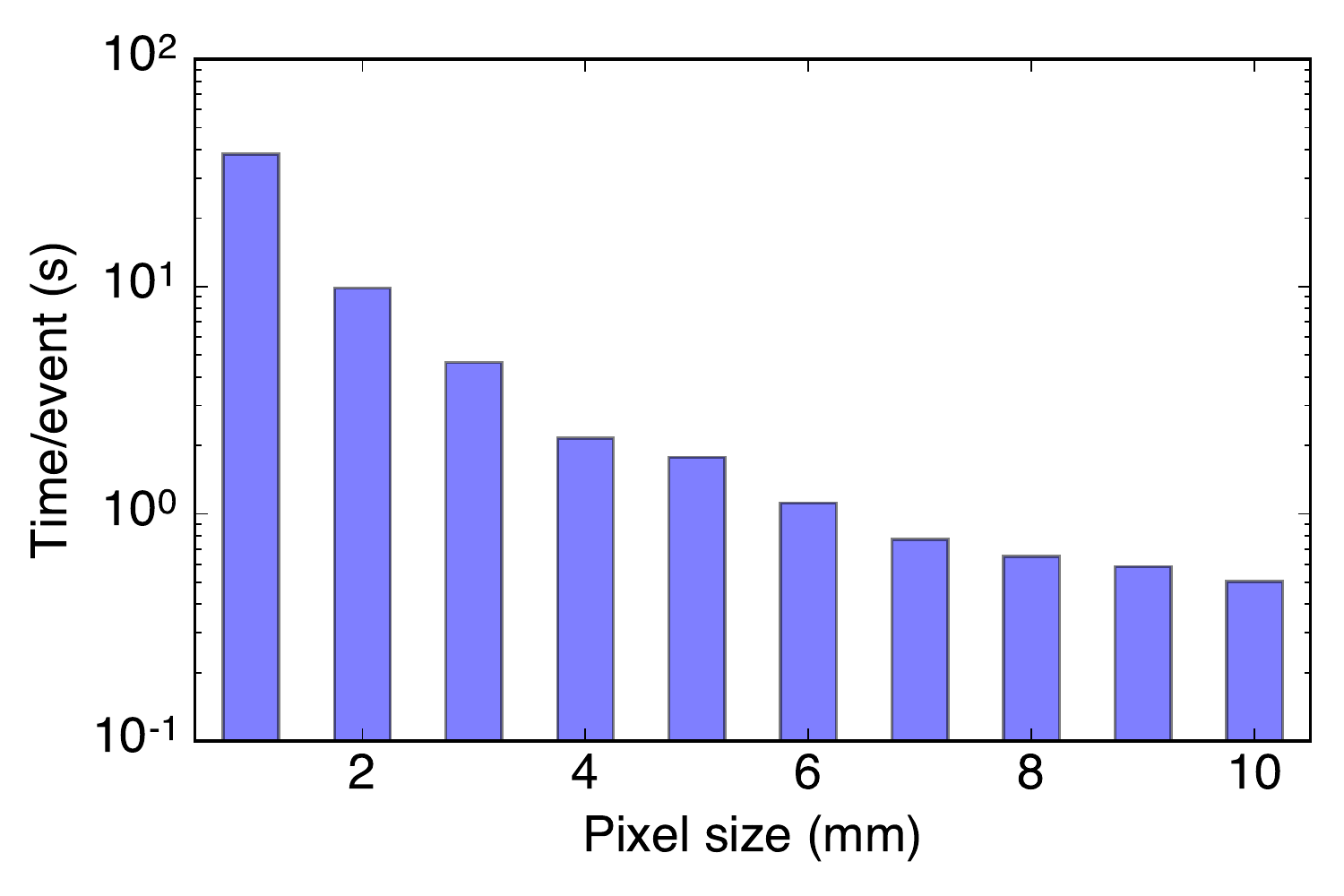}
		\caption{\label{fig:timeSize} Computing time per event for different pixel sizes after 100 iterations.}
\end{figure}

Computing time evaluation of the choice of pixel size shows that time is increased dramatically as the image pixel used is smaller (fig.\ \ref{fig:timeSize}). Therefore, since performance of the method does not improves significantly with small pixels, a 10 mm pixel size is desirable at least for a fast event selection and reconstruction while detailed reconstruction should be done only for the events within the region of interest.

\subsection{Radial dependence}
\label{sec:perform:radial}

The proximity of the event to the field cage walls can affect the performance of the reconstruction since near these walls the border effects impact the signal more significantly. An evaluation of how these effects worsens the performance of our implementation of the method and chosen model is then needed. For this, several radial fiducial cuts (from 50 mm to 200 mm in 10 mm jumps) are applied to krypton events reconstructed using ML-EM after applying 100 iterations using a pixel size of 10 mm. The fiducial cut is done by demanding that the reconstructed mean position of the event is within the fiducial area.

\begin{figure}[htbp]
	\centering 
	\includegraphics[width=0.495\textwidth]{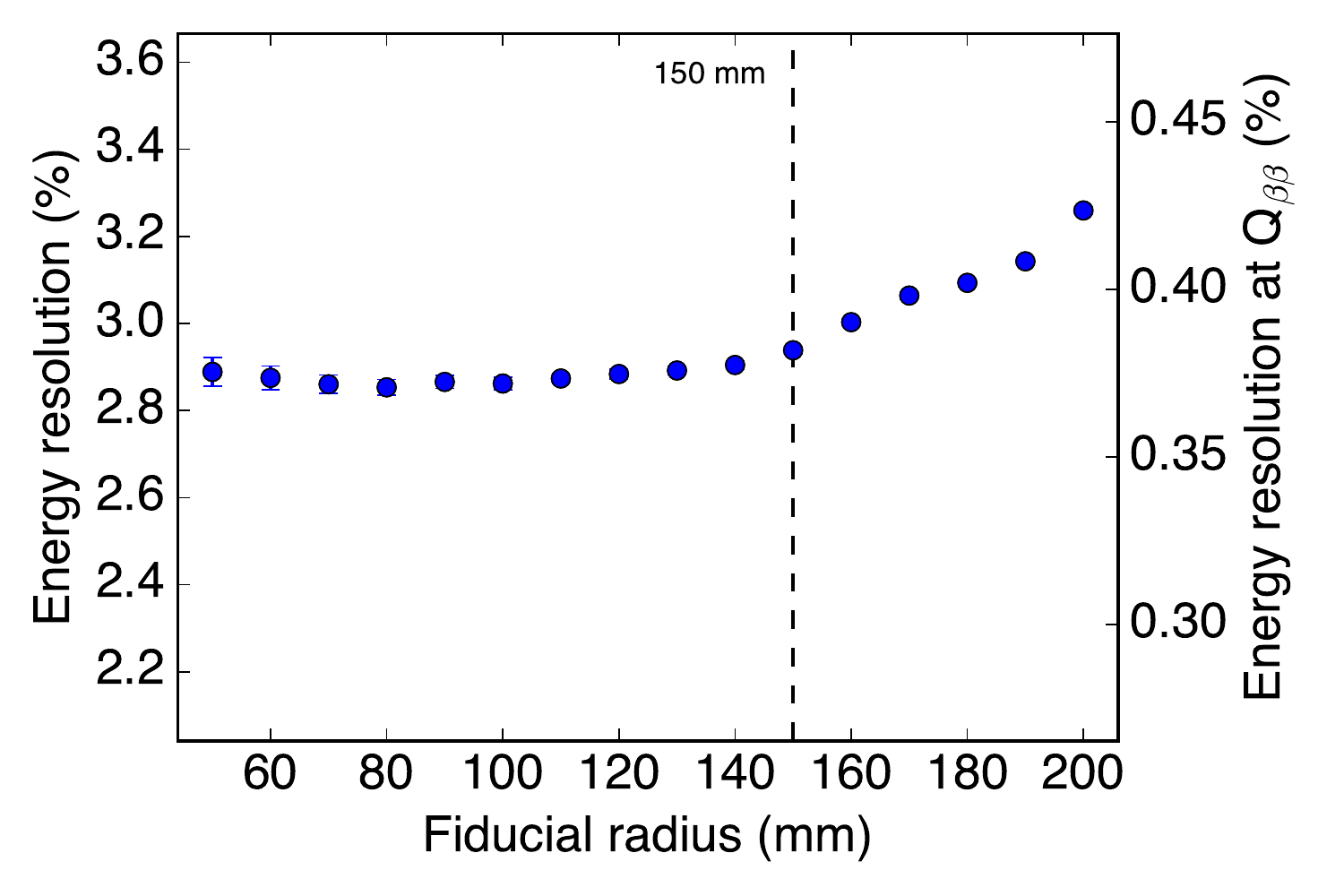}
	\includegraphics[width=0.495\textwidth]{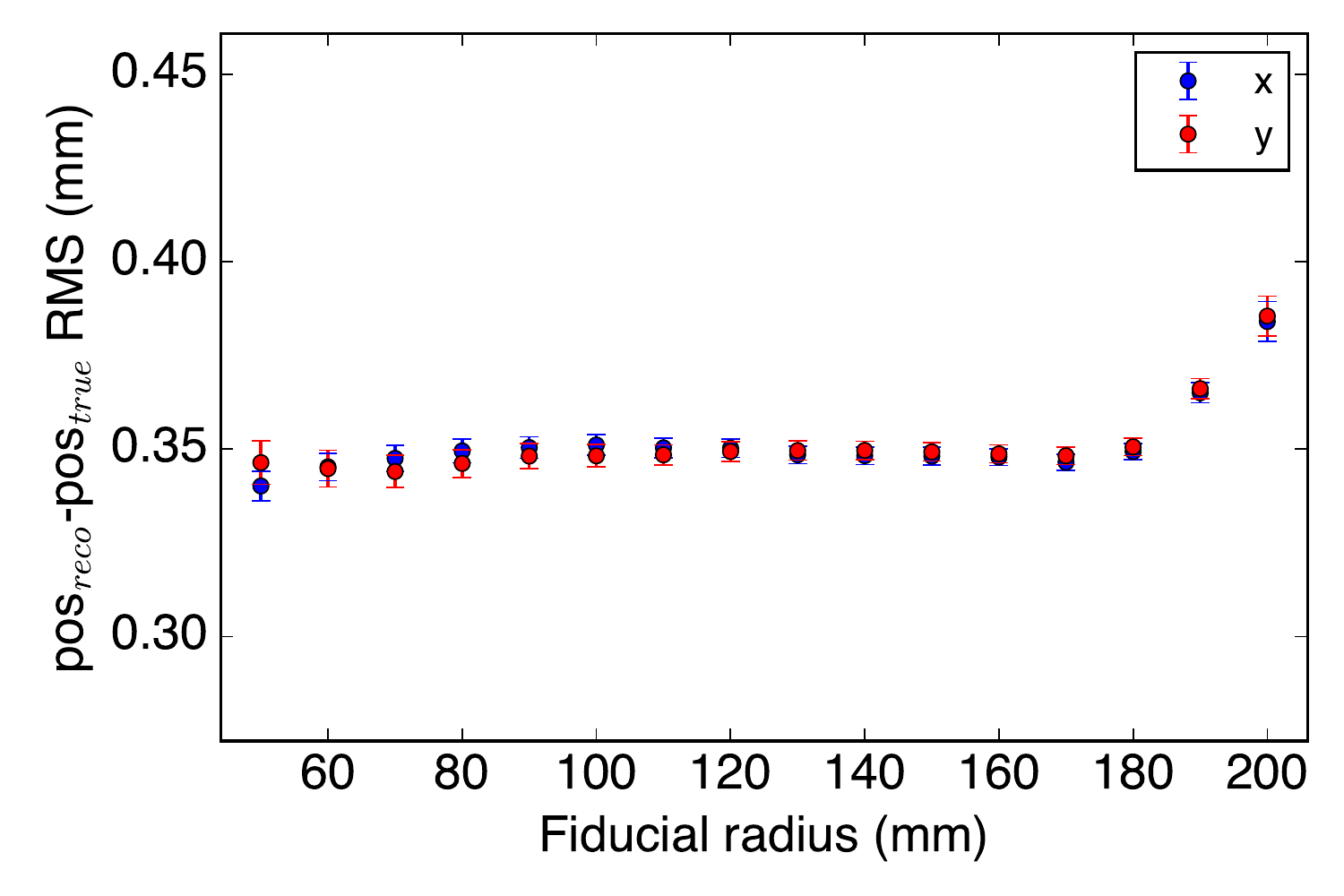}
	\caption{\label{fig:Rad} Dependence on the fiducial radius of energy resolution (left) and of the rms of the XY mean position difference between reconstructed and true information (right) for $^{83}$Kr.}
\end{figure}

Energy resolution remains stable up to a radial distance of 150 mm from the center ($\sim$2.9$\%$ FWHM) (fig.\ \ref{fig:Rad}, left). From that point resolution begins to get worse to end up at $\sim$ 3.25$\%$ FWHM for the whole active volume of the detector. Although resolution deteriorates by 10$\%$, resolution for the full active volume is better than the Collaboration's objective of 0.5 $\%$ at the $Q_{\beta\beta}$ value.

On the other hand, the difference between true and reconstructed position remains stable up to a radial distance of 180 mm at a value of $\sim$ 0.35 mm for each dimension (fig.\ \ref{fig:Rad}, right). After that the difference increases but still remains below 0.4 mm for the full active volume of the chamber. This could be due to the complicated parametrization that is used.

Considering both parameters, best results are achieved for events contained within 150 mm from the longitudinal axis of  the chamber. However, since the number of events increases with the square of the radius being able to use the algorithm successfully with large fiducial areas can be of potential utility at the cost of having a slightly worse performance.

\subsection{Cut over pixel charge}
\label{sec:perform:cut}

When creating the seed, pixels which could have charge according to the light detected are taken into account. However, this does not imply that all the pixels considered actually have charge. The inclusion of non-relevant pixels then can add unnecessary extra-time to the method. To avoid this a cut over the charge of the pixels can be done at each iteration. The cut consists on assigning a charge of zero to pixels below a threshold equal to a percentage of the highest charge of all pixels. 

\begin{figure}[htbp]
	\centering 
	\includegraphics[width=0.495\textwidth]{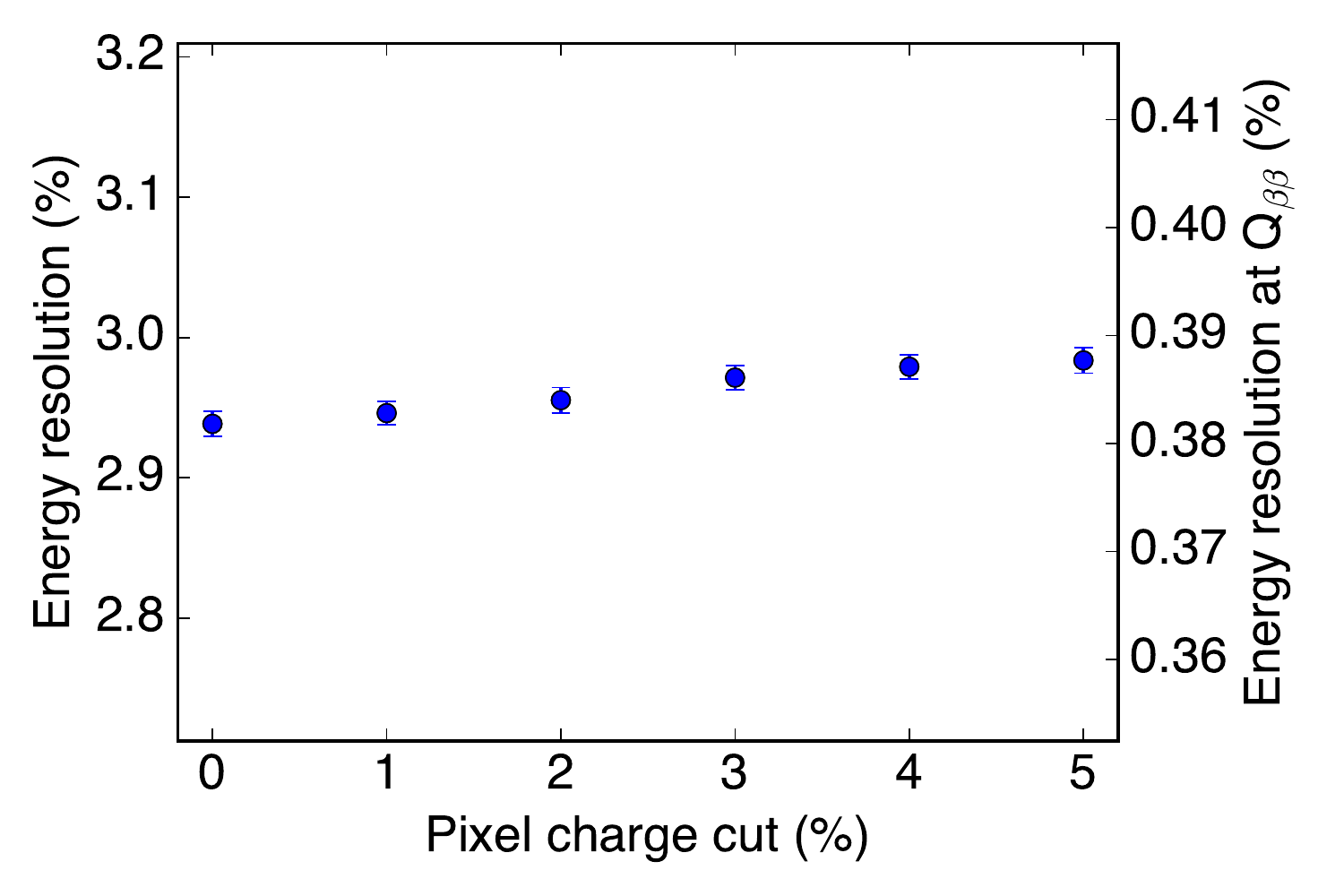}
	\includegraphics[width=0.495\textwidth]{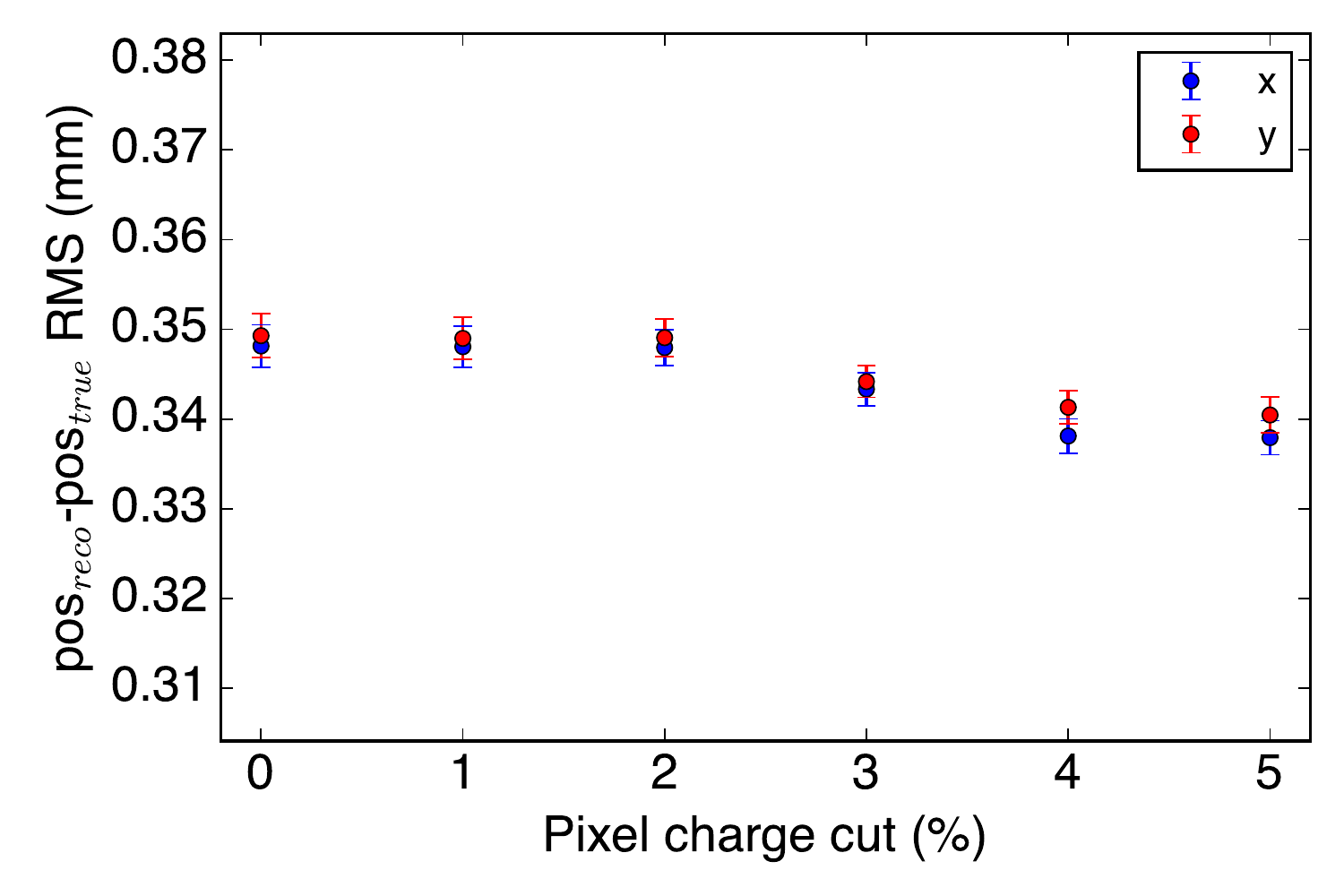}
	\caption{\label{fig:Cut} Pixel charge cut dependence of the energy resolution (left) and of the rms of the XY mean position difference between reconstructed and true information (right) for $^{83}$Kr.}
\end{figure}

The method has been applied through 100 iterations with a pixel size of 10 mm and a fiducial radius of 150 mm. The pixel charge cuts evaluated range from 1$\%$ to 5$\%$ in 1$\%$ jumps. Energy resolution gets slightly worse as we increase the threshold for the cut but remains within acceptable values, between 2.94$\%$ and 2.99$\%$ FWHM at the 41.5 keV of the krypton event (fig.\ \ref{fig:Cut}, left).

In contrast, the difference in position shows an opposite behaviour (fig.\ \ref{fig:Cut}, right). When increasing the cut and, therefore, reducing the total number of pixels, the available charge is concentrated around the real position of the event reducing the difference between the reconstructed position and the real one. Still, as was the case with the energy resolution, the performance regarding this parameter is not greatly affected by the cut values as the difference between the best and the worst case is only $\sim$0.1 mm.

\begin{figure}[htbp]
		\centering
		\includegraphics[width=0.6\textwidth]{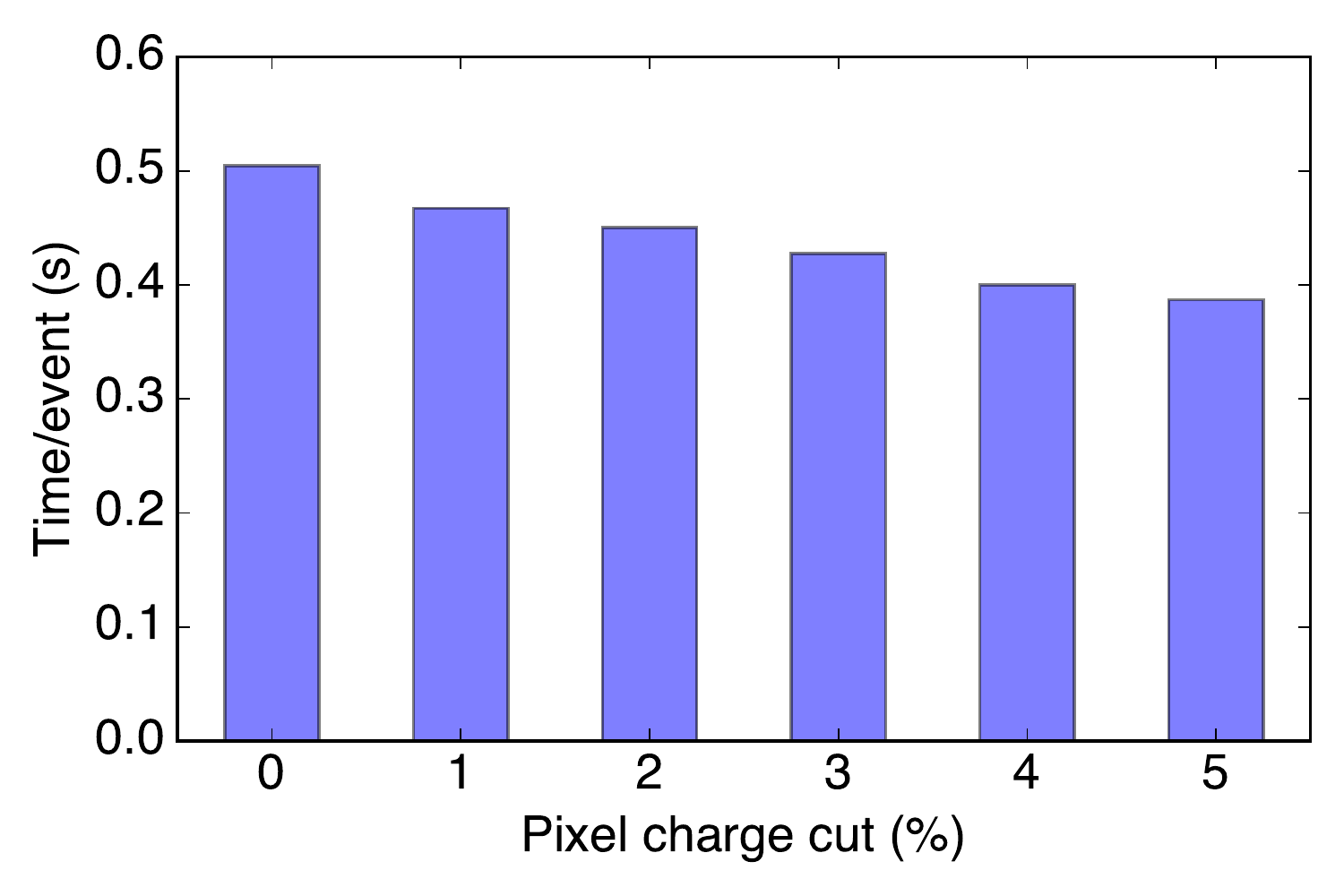}
		\caption{\label{fig:timeCut} Computing time per event for different pixel charge cuts.}
\end{figure}

Although reducing it, the pixel charge cut does not impact significantly on computing time of the method (fig.\ \ref{fig:timeCut}). A better reconstruction performance is then prioritized over time performance and, therefore, the cut should be avoided.

\subsection{Energy dependence}
\label{sec:perform:energy}

All the evaluations so far have been made considering point-like depositions. However, most of the events in consideration in NEXT will be higher energy events producing long tracks within the detector. The difference in mean position is not that relevant for this kind of event and the validity of tracking reconstruction in these cases should be evaluated using another approach and will be a focus of future work.

However, a correct extrapolation and determination of the energy resolution for higher energies is absolutely necessary. To evaluate the performance of the method in this regard, simulated events of the 511 keV and 1.2 MeV gammas from $^{22}$Na, 1.6 MeV electron-positron pair from $^{208}$Tl and $\beta\beta 0 \nu$ events of $^{136}$Xe (2.458 MeV) have been reconstructed with the ML-EM algorithm using 10$times$10 mm pixels and 100 iterations. This has been done for events within a  radial distance of $150$ mm from the chamber's longitudinal axis.

\begin{figure}[htbp]
	\centering 
	\includegraphics[width=0.495\textwidth]{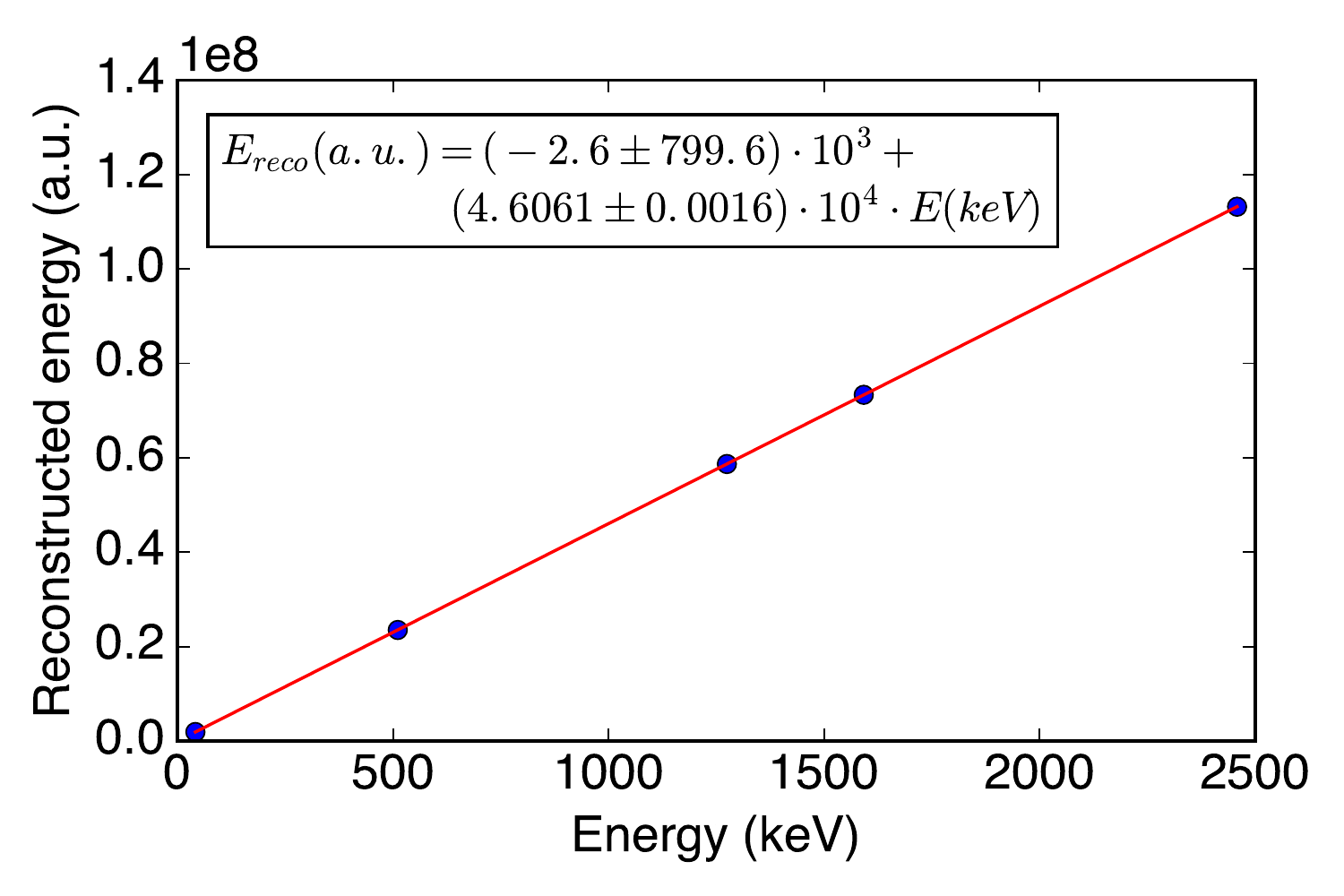}
	\includegraphics[width=0.495\textwidth]{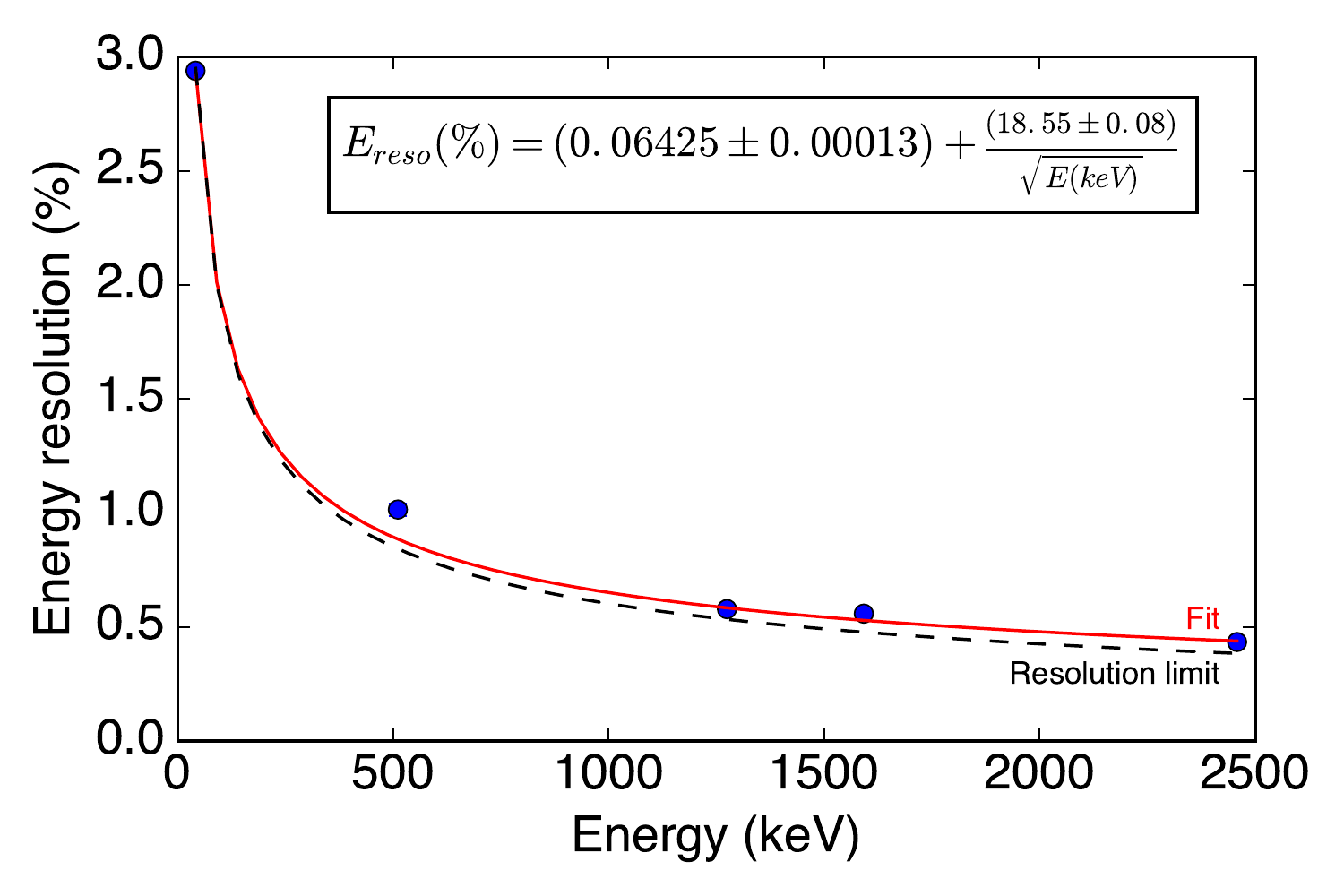}
	\caption{\label{fig:Energy} Scalability of the reconstructed energy (left) and of the energy resolution for a fiducial cut of 150 mm (right) for $^{83}$Kr, $^{22}$Na, $^{208}$Tl and $^{136}$Xe $\beta\beta 0 \nu$ events. In red a fit to the results (blue) from ML-EM reconstruction; the dashed line shows the energy resolution limit in gaseous xenon under the simulated parameters.}
\end{figure}

The reconstructed energy for the various sources shows a linear behaviour (fig.\ \ref{fig:Energy}, left) allowing for an easy calibration method when applying it to real data. On the other hand, energy resolution scales inversely with the square root of the event energy as it is expected if there are no other effects dominating the resolution (fig.\ \ref{fig:Energy}, right). 

The energy resolution limit of a gaseous detector is described by: 

\begin{equation}
	\delta E/E = 2.35\sqrt{\frac{W_i(F+G)}{E}},
	\label{eq:resolimit}
\end{equation}
where $W_i$ is the average energy needed to create an electron-ion pair, $F$ is the Fano Factor and accounts for the fluctuations in the number of ionizations produced, $G$ describes the variations derived from the detection process and $E$ is the total energy deposited in the gas.

For gaseous xenon, $W_i$ is estimated to be $22.4~eV$ \cite{cite:ionE}, $F$ is 0.15 \cite{cite:fano} and G can be simplified \cite{cite:factorG}, when working at the gain considered in the simulations, to:

\begin{equation}
	G = \frac{(\sigma_q/q)^2 + 1}{\eta Y},
	\label{eq:factorG}
\end{equation}
where $\sigma_q/q$ is the charge resolution of the PMTs (0.6 in the simulation) and $\eta Y$ is the number of photons detected per ionization electron (9.49 according to the detection probabilities considered). Considering all, the energy resolution limit would be $\sim0.384\%$ FWHM at $Q_{\beta\beta}$.

The results obtained at all energies match almost completely the best resolution attainable with the simulation parameters used for data generation. It follows that, with the appropriate model, the method will perform outstandingly.

\begin{figure}[htbp]
	\centering 
	\includegraphics[width=0.45\textwidth]{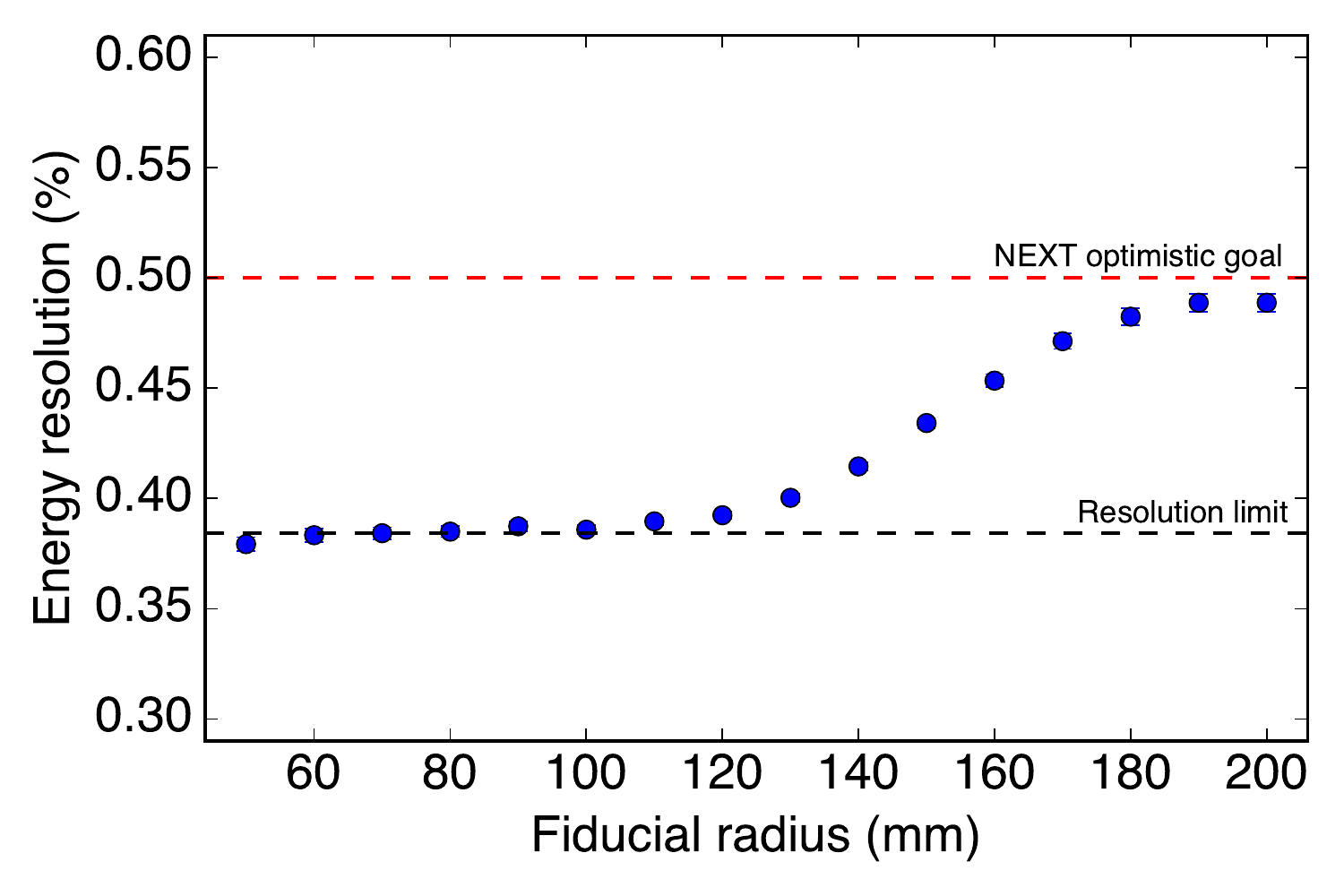}
	\includegraphics[width=0.5\textwidth]{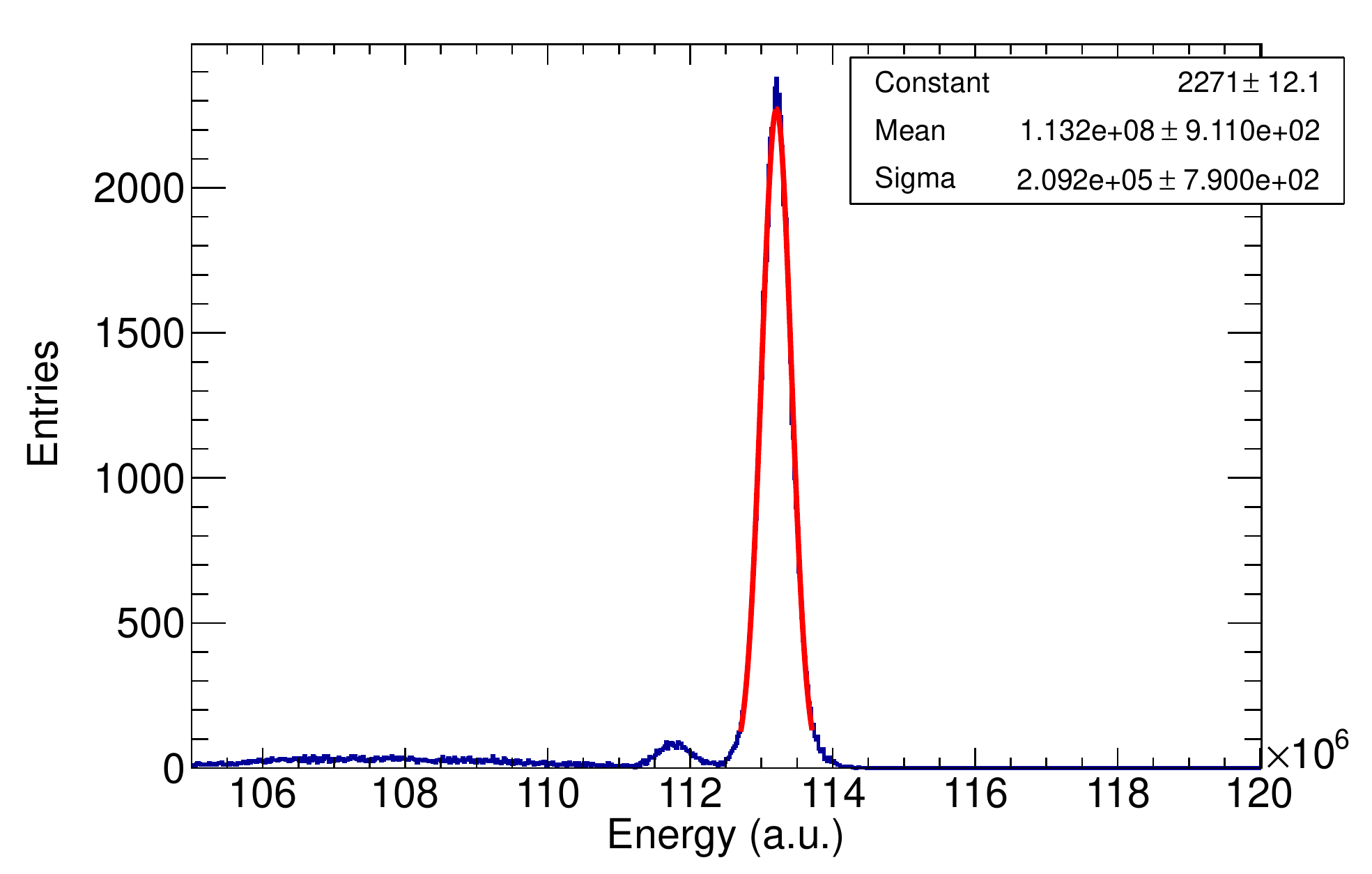}
	\caption{\label{fig:resoQbb} Radial dependence of the energy resolution for $^{136}$ Xe $\beta \beta 0 \nu$  events (left). Resolution for a 150 mm fiducial cut shows a 0.434\% FWHM energy resolution (right).}
\end{figure}

Energy resolution radial dependence has already been evaluated with point-like depositions. However, it could have a different impact on the case of long tracks. Evaluating this effect in the same way as in section \ref{sec:perform:radial} for $^{136}$Xe $\beta \beta 0 \nu$ events shows that indeed the resolution gets a bit worse when considering bigger volumes but even when considering the full volume of the chamber, resolution of neutrinoless double beta decays is of 0.49\% FWHM (fig.\ \ref{fig:resoQbb}, left). 

However, fiducial cuts enhance even more the energy resolution. Concretely, energy resolution for a reasonable fiducial radial cut up to 150 mm, when resolution starts to deteriorate, is 0.43\% FWHM. If the fiducial volume is reduced enough, resolution achieved reaches the limit derived from physics itself. 

\subsection{Track reconstruction}
\label{sec:perform:track}

As stated, ML-EM not only allows for a great energy resolution but also provides a detailed reconstruction of the event. This characteristic is vital for background rejection in NEXT since the topological cut is especially relevant. 

\begin{figure}[htbp]
	\centering 
	\includegraphics[width=0.44\textwidth]{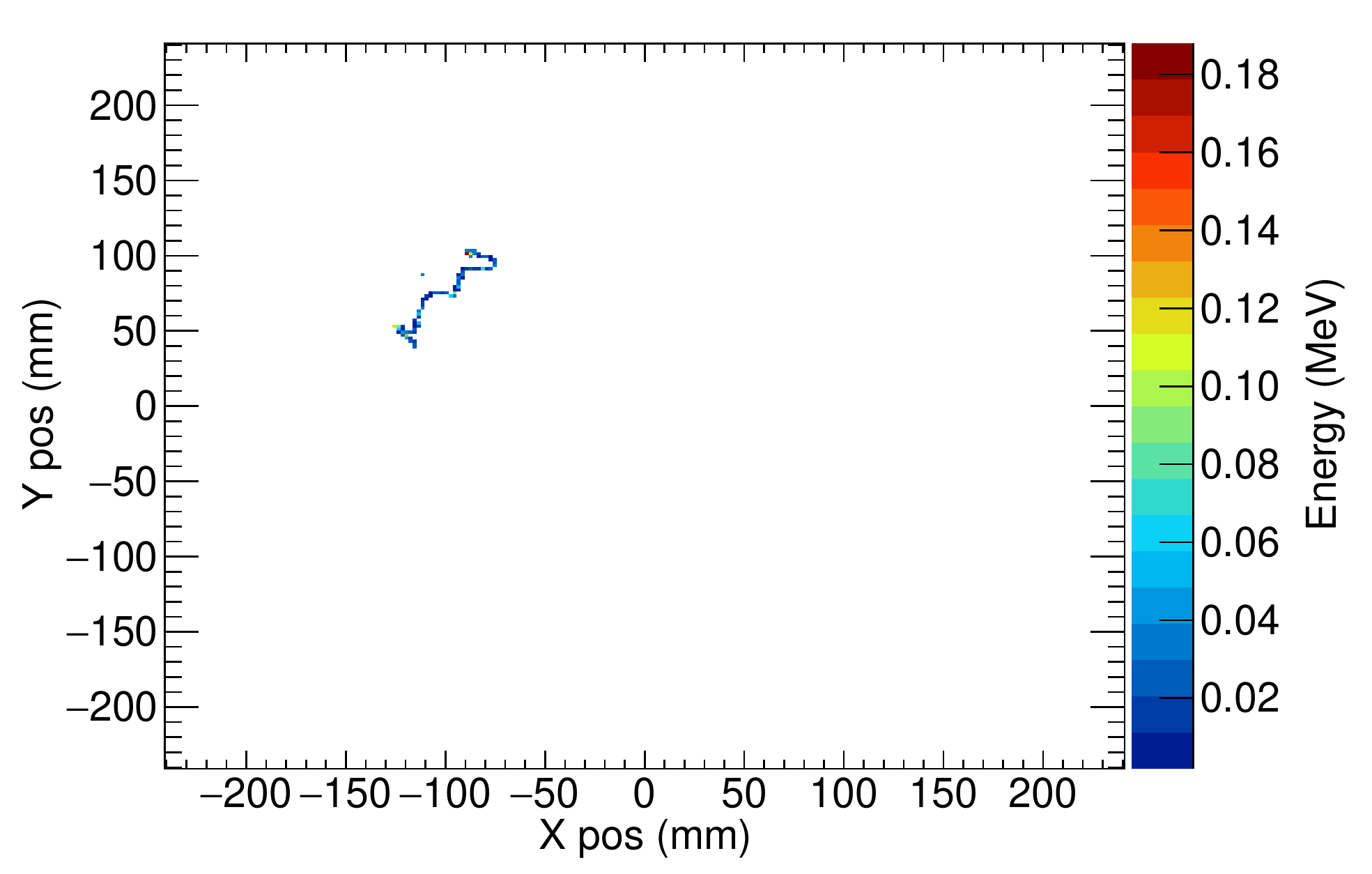}
	\includegraphics[width=0.44\textwidth]{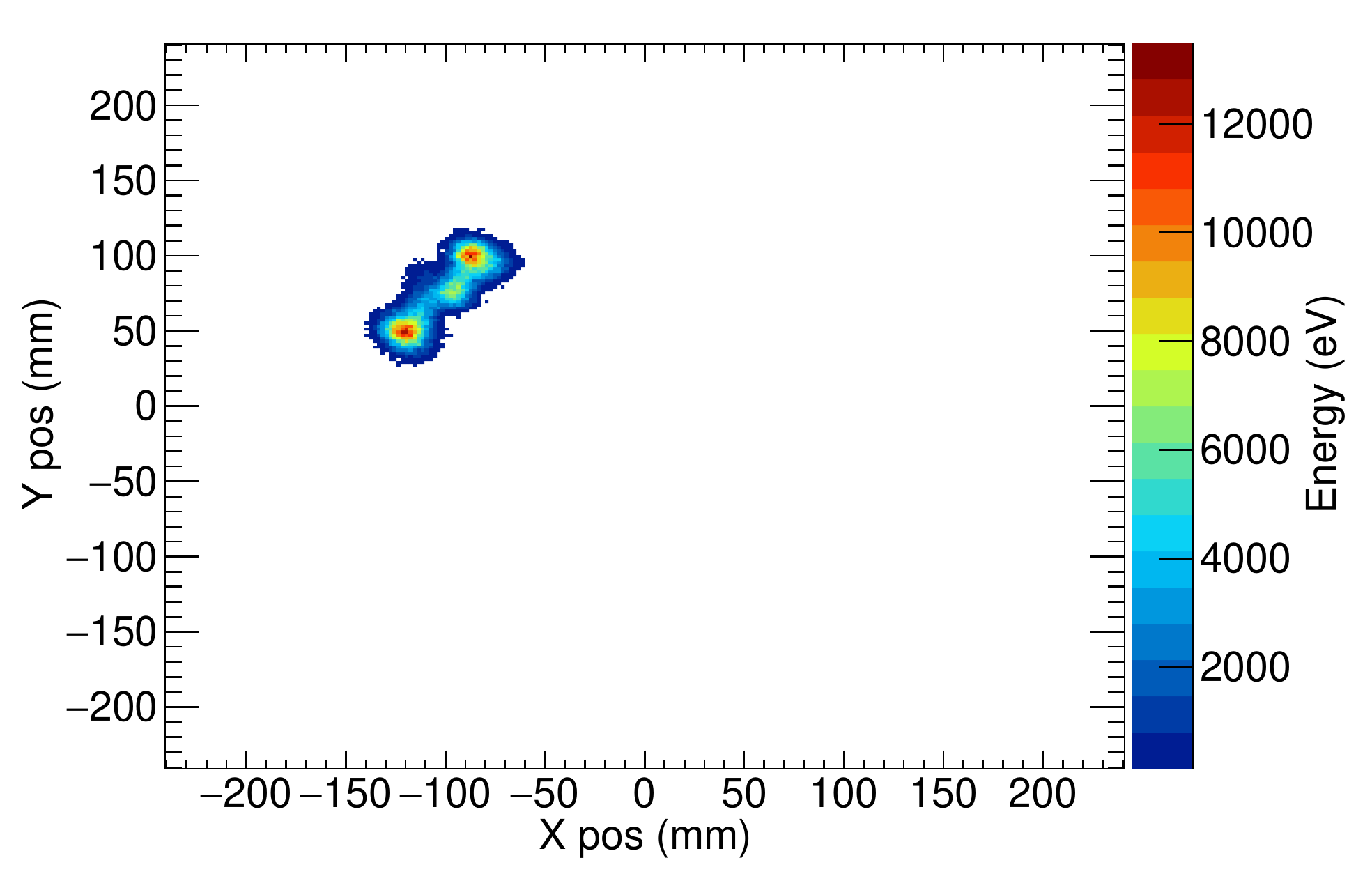}
	\includegraphics[width=0.44\textwidth]{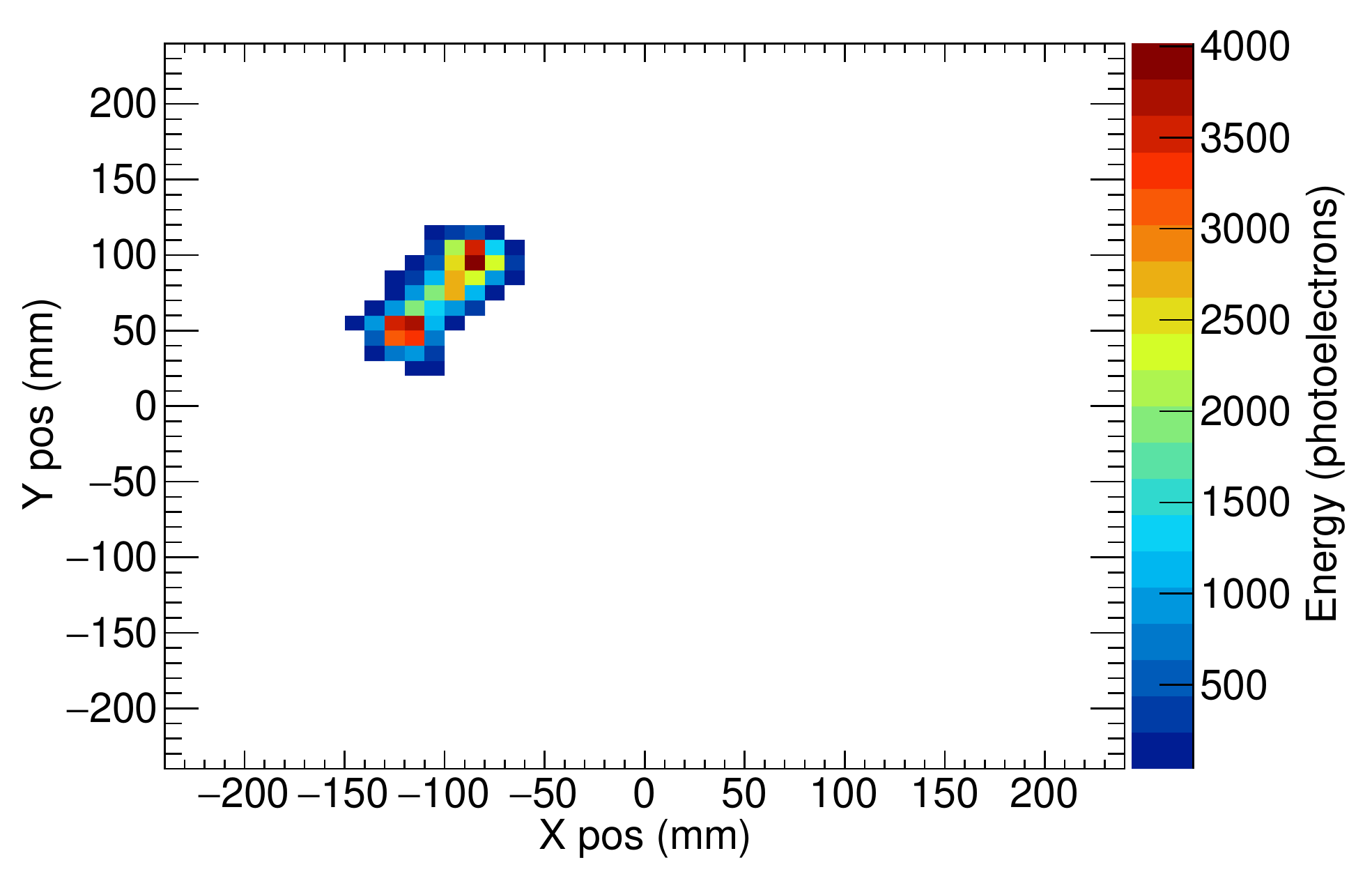}
	\includegraphics[width=0.44\textwidth]{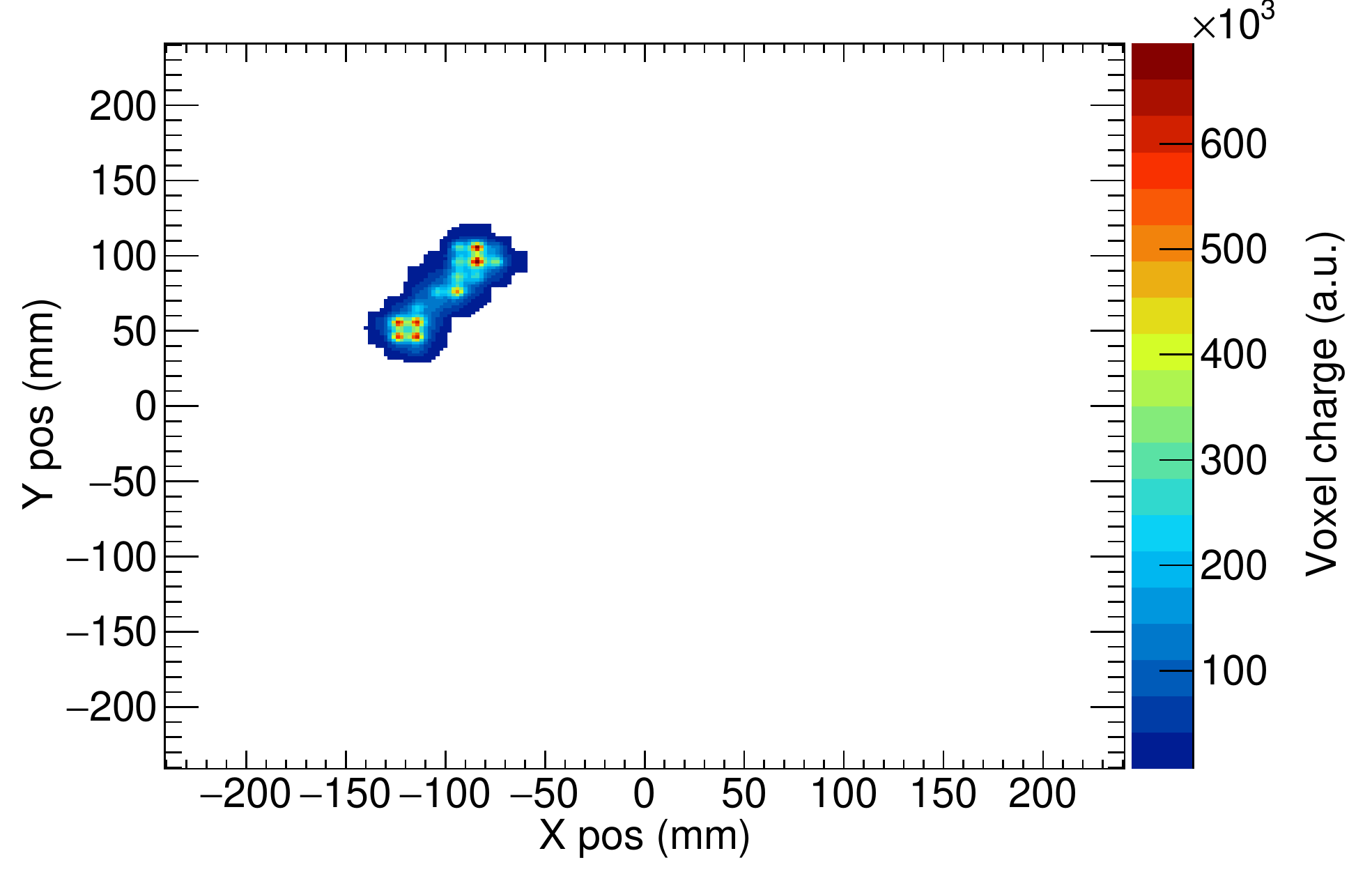}
	\caption{\label{fig:track} Simulated $\beta\beta 0 \nu$ track in NEW (top left) and the arriving charge produced by it that arrives to the electroluminescence region (top right). Anode response to the light produced of this charge (bottom left) and reconstructed track using ML-EM with 100 iterations (bottom right). Since diffusion is not being deconvolved, pictures on the right should be compared.}
\end{figure}

A $\beta\beta 0 \nu$ track (fig.\ \ref{fig:track}, top) reconstructed using 100 iterations of the two-dimensional mode of the ML-EM algorithm with a pixel size of 2 mm is shown on Figure \ref{fig:track}, bottom right. It can be seen that with these parameters the method tends to concentrate the charge in certain spots instead of providing a smooth distribution. The position of the hot spots coincide with the center of the SiPMs located in the anode, where the probability of detection is higher.

\begin{figure}[htbp]
	\centering 
	\includegraphics[width=0.495\textwidth]{imgs/difftrack.pdf}
	\includegraphics[width=0.495\textwidth]{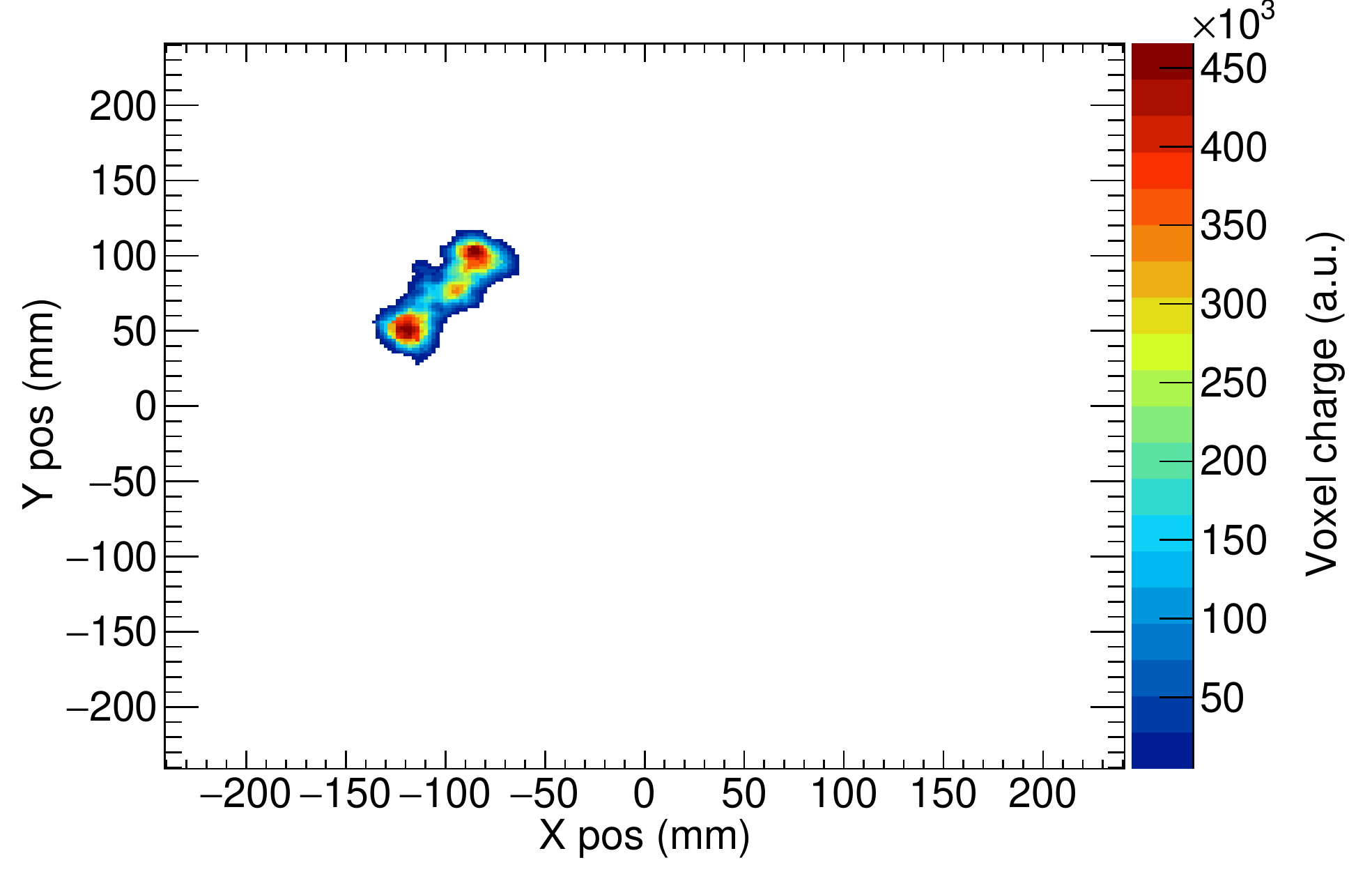}
	\caption{\label{fig:trackReco1000} Simulated $\beta\beta 0 \nu$ track smeared by diffusion (left) and its reconstruction using 1000 iterations of ML-EM (right).}
\end{figure}

However this quantization effect vanishes for higher number of iterations and it can be seen that with 1000 iterations the track is perfectly reconstructed (fig.\ \ref{fig:trackReco1000}, right). From this it can be concluded that the likelihood cut selection of 0.1\% is enough for precise energy reconstruction but not for track recovery. 

It should be noted that at its current state, the diffusion probability is not implemented and the tracks reconstructed are smeared by diffusion. Therefore the reconstructed track should be compared with the diffused simulated track (fig.\ \ref{fig:track}, top right). Both the reconstructed and the simulated track share the same traits and the characteristic two blobs of a $\beta\beta 0 \nu$ event in gas are clearly seen. 

\section{Conclusions and future work}
\label{sec:future}

Application of the ML-EM algorithm within the reconstruction scheme of NEXT has shown an outstanding performance so far. Evaluation of the method shows that an energy resolution of (0.434 $\pm$ 0.002)\% FWHM (error derived from the fit to the energy spectrum) and precise event positioning (within 1.7 mm from the real position with 3$\sigma$) can be achieved with only 100 iterations of the method for most of the detector active volume (fiducial radius of 150 mm). When considering the full active volume then resolution becomes (0.489 $\pm$ 0.003)\%. This performance surpasses the optimistic goal of the Collaboration of a 0.5$\%$ energy resolution for $\beta\beta 0 \nu$ events. For precise tracking reconstruction around 1000 iterations are needed alongside a new convergence criteria. However, a fast selection of events within the energy region of interest can be easily made thanks to the fast convergence of the algorithm when considering energy resolution.

Implementation of tri-dimensional reconstruction modes is currently undergoing as well as the inclusion of diffusion into the probability model. While energy resolution will not, in principle, change since the total charge reconstructed should be the same, an improvement in the tracking reconstruction should be expected. 

Other effects, as those derived from the drift in electric fields \cite{cite:CondeField} will be evaluated in future works. In addition, a detailed study of the impact of the method on background discrimination will be done.


\acknowledgments

The NEXT Collaboration acknowledges support from the following agencies and institutions: the European Research Council (ERC) under the Advanced Grant 339787-NEXT; the Ministerio de Econom\'ia y Competitividad of Spain under grants FIS2014-53371-C04 and the Severo Ochoa Program SEV-2014-0398; the GVA of Spain under grant PROMETEO/2016/120; the Portuguese FCT and FEDER through the program COMPETE, project PTDC/FIS/103860/2008; the U.S.\ Department of Energy under contracts number DE-AC02-07CH11359 (Fermi National Accelerator Laboratory) and DE-FG02-13ER42020 (Texas A\&M); and the University of Texas at Arlington. 


\end{document}